\documentclass[12pt]{article}
\usepackage{graphicx}% Include figure files
\usepackage{bm}% bold math
\def\be{\begin{equation}}
\def\ee{\end{equation}}
\def\bea{\begin{eqnarray}}
\def\eea{\end{eqnarray}}
\usepackage{latexsym}
\usepackage{dcolumn}
\usepackage{epsfig,amssymb,euscript}
\usepackage{amsmath}
\usepackage{array,calc,epsfig}
\DeclareMathOperator\arctanh{arctanh}

\usepackage{chngcntr}
\usepackage[noadjust]{cite}

\usepackage{color}
\renewcommand{\to}{\rightarrow}
\renewcommand{\l}{\lambda}

\def\be{\begin{equation}}
\def\ee{\end{equation}}
\def\ba{\begin{eqnarray}}
\def\ea{\end{eqnarray}}
\def\nb{\nonumber}
\def\p{\partial}

\def\a{\alpha}
\def\b{\beta}

\def\ff{\phi}

\def\l{\lambda}
\def\L{\Lambda}

\def\o{\omega}
\def\O{\Omega}
\def\r{\rho}
\def\s{\sigma}

\def\th{\theta}

\def\q{\quad}

\def\mc{\mathcal}

\def\ra{\rightarrow}
\def\lra{\leftrightarrow}

\newcommand{\pr}[1]{\left(#1\right)}
\newcommand{\pq}[1]{\left[#1\right]}

\counterwithin*{equation}{section}

\begin{document}
\baselineskip=15.5pt
\pagestyle{plain}
\setcounter{page}{1}
\newfont{\namefont}{cmr10}
\newfont{\addfont}{cmti7 scaled 1440}
\newfont{\boldmathfont}{cmbx10}
\newfont{\headfontb}{cmbx10 scaled 1728}
\renewcommand{\theequation}{{\rm\thesection.\arabic{equation}}}
%\font\cmss=cmss10 \font\cmsss=cmss10 at 7pt
\renewcommand{\thefootnote}{\arabic{footnote}}

\vspace{1cm}
\begin{titlepage}
\vskip 2cm
\begin{center}
{\Large{\bf Fate of false vacua in holographic \\ first-order phase transitions}}
\end{center}

\vskip 10pt
\begin{center}
Francesco Bigazzi$^{a}$, Alessio Caddeo$^{a,b,}$\footnote{On leave at the Universit\'e Libre de Bruxelles; C.P. 231, 1050 Brussels, Belgium.}, Aldo L. Cotrone$^{a,b,}$\footnote{On leave at the Galileo Galilei Institute for Theoretical Physics, INFN National Center for Advanced
Studies, Largo E. Fermi, 2, 50125 Firenze, Italy.}, Angel Paredes$^{c}$
\end{center}
\vskip 10pt
\begin{center}
\vspace{0.2cm}
\textit {$^a$ INFN, Sezione di Firenze; Via G. Sansone 1; I-50019 Sesto Fiorentino (Firenze), Italy.
}\\
\textit{$^b$ Dipartimento di Fisica e Astronomia, Universit\'a di Firenze; Via G. Sansone 1;\\ I-50019 Sesto Fiorentino (Firenze), Italy.
}\\
\textit{$^c$ Departamento  de  Fisica  Aplicada,  Universidade  de  Vigo,  As  Lagoas  s/n,  Ourense,  ES-32004  Spain.}\\
\vskip 20pt
{\small{
bigazzi@fi.infn.it, alessio.caddeo@unifi.it, cotrone@fi.infn.it, angel.paredes@uvigo.es
}

}

\end{center}

\vspace{25pt}

\begin{center}
 \textbf{Abstract}
\end{center}

\noindent 
Using the holographic correspondence as a tool, we study the dynamics of first-order phase transitions in 
strongly coupled gauge theories at finite temperature. Considering an evolution from the large to the small temperature phase, we compute the nucleation rate of bubbles of true vacuum in the metastable phase. For this purpose, we find the relevant configurations (bounces) interpolating between the vacua and we compute the related effective actions. We start by revisiting the compact Randall-Sundrum model at high temperature. Using holographic renormalization, we compute the kinetic term in the effective bounce action, that was missing in the literature. Then, we address the full problem within the top-down Witten-Sakai-Sugimoto model. It displays both a confinement/deconfinement and a chiral symmetry breaking/restoration phase transition which, depending on the model parameters, can happen at different critical temperatures. For the confinement/deconfinement case we perform the numerical analysis of an effective description of the transition
and also provide analytic expressions using thick and thin wall approximations. For the chiral symmetry transition, we implement a variational approach that allows us to address the challenging non-linear problem stemming from the Dirac-Born-Infeld action. 

\end{titlepage}
\newpage
\tableofcontents

\section{Introduction}

The gauge/gravity duality provides unique tools to study the properties of strongly coupled gauge theories, including their phase structure.
First-order phase transitions have been thoroughly analyzed in many different models,
following the seminal papers \cite{Witten:1998qj} for theories with only adjoint matter, and \cite{Mateos:2006nu} for cases with fundamental matter. 
Once the threshold for the phase transition is crossed, the former minimal energy configuration becomes a ``false vacuum'' and is expected to
decay to the new ground state, the ``true vacuum''.

This kind of vacuum decay was first studied long ago in a simple one-scalar field model  \cite{Coleman:1977py}, where a first-order phase transition occurs when the scalar potential has two minima, one of which is metastable.
The decay of the latter can proceed through quantum tunneling or via thermal fluctuations (or, more generally, by a combination of the two effects). Dynamically, the transition happens through the nucleation of bubbles of true vacuum in the metastable phase \cite{Coleman:1977py,Callan:1977pt,Coleman:1980aw,Linde:1980tt,Linde:1981zj}.

In general, the decay rate of a metastable vacuum per unit volume in the semiclassical limit is given by an expression of the form $\Gamma= A\, e^{-S_B }$, where $A$ and $S_B$ depend on the underlying quantum field theory. The first coefficient is usually very hard to compute in closed form: it is given in terms of a certain functional determinant and it is often estimated by dimensional analysis. The exponential term is the so-called {\it bounce} action. For a scalar field $\Phi$ in 3+1 dimensions, with potential having an absolute minimum (the true vacuum) at $\Phi_t$ and a local minimum (the false vacuum) at $\Phi_f$, the bounce action is defined by $S_B  = S_E(\Phi_B) - S_E(\Phi_f)$, where $S_E$ is the Euclidean action for the scalar field and $\Phi_B$ is called ``the bounce". The latter is a non-trivial ``bubble-like'' solution of the Euclidean equation of motion which approaches the false vacuum $\Phi_f$ at Euclidean infinity and a constant $\Phi _0$ at the center of the bubble.\footnote{As discussed in \cite{Coleman:1977py,Coleman:1980aw}, this Euclidean solution is meant to represent the bubble at time zero in Minkowskian signature.}
When the transition proceeds through quantum tunneling, the bounce is $O(4)$ symmetric and $\Phi_B$ only depends on the radial coordinate $\rho=\sqrt{t^2+ x_i x_i}$, where $t$ is the Euclidean time and $x_i$ are the space coordinates. When the transition is dominated by thermal fluctuations, the bounce is $O(3)$ symmetric and $\Phi_B=\Phi_B(\rho)$, with $\rho=\sqrt{x_i x_i}$.  The configuration for which the rate $\Gamma$ has the larger value is the one that dominates the decay process. 

The main aspects of this simple scalar model can be generalized to vacuum decay in gravitational dual descriptions of quantum field theories, a process that has been studied in various papers in the past. Nevertheless, as far as we know, this literature is focused on bottom-up models with AdS geometries, like those relevant for Randall-Sundrum (RS)-like setups \cite{Creminelli:2001th,Randall:2006py,Nardini:2007me,Konstandin:2010cd,Bunk:2017fic,Dillon:2017ctw,Megias:2018sxv,Baratella:2018pxi,Agashe:2019lhy,Fujikura:2019oyi,DelleRose:2019pgi,vonHarling:2019gme,Megias:2020vek}.\footnote{See \cite{Horowitz:2007fe} for some considerations on backgrounds dual to confining theories.} 

In this paper we try to proceed a step further, studying, for the first time, the dynamics of first-order phase transitions in gauge theories with a precise string embedding. This top-down perspective allows for a precise identification of the gauge theories under investigation and for an understanding of the approximations leading to the dual classical gravitational descriptions. As a result, computations performed in the planar limit at strong coupling are reliable, without uncontrolled approximations as the ones plaguing effective models (such as sigma models, NJL, etc.) or bottom-up holographic models.

The theories we focus on are based on the Witten-Sakai-Sugimoto (WSS) model  \cite{Witten:1998zw,Sakai:2004cn}.
It is the top-down holographic theory closest to QCD and it has been very successful in modeling aspects of its strong coupling dynamics.
In the limits where a simple dual classical gravitational description is available, the model consists of a large $N$, $SU(N)$ gauge theory coupled to $N_f\ll N$ fundamental fermions and to a tower of adjoint massive Kaluza-Klein (KK) matter fields. The latter arise from the fact that the Yang-Mills sector of the theory describes the low energy dynamics of a stack of $N$ D4-branes wrapped on a circle of coordinate $x_4\sim x_4 + 2\pi/M_{KK}$. The fundamental chiral fields in the model are added by means of further $N_f$ D8/anti-D8 (``flavor'') brane pairs, placed at fixed points on the above-mentioned circle, asymptotically separated by a distance $L$. In the $N_f\ll N$ limit, the backreaction of the flavor branes on the dual gravity background can be neglected.\footnote{Going beyond this leading order quenched regime in WSS is indeed possible. See \cite{Burrington:2007qd} and \cite{Bigazzi:2014qsa} for related results.} 

The WSS model exhibits two kinds of first-order phase transitions at finite temperature. The confinement/deconfinement phase transition occurs at a critical temperature $T_c= M_{KK}/2\pi $. In the dual gravity picture, it corresponds to a Hawking-Page transition between a solitonic background and a black brane solution. 
Finding the full-fledged  configuration that interpolates between the two backgrounds in ten-dimensional supergravity is an extremely interesting but complicated open problem, see e.g. \cite{Aharony:2005bm}. Following a prescription developed in bottom-up RS-AdS models in \cite{Creminelli:2001th},
we will use an off-shell description of the phase transition, modeling it with a single scalar effective action which 
we will compute using holographic renormalization techniques. 
From this, we will compute 
the aforementioned bounce, effectively interpolating between the two vacua, and its on-shell action.
This will allow us to determine the bubble nucleation rate in terms of the parameters of the model.

If the flavor branes are placed at antipodal points on the compactification circle, i.e. when $L M_{KK}=\pi$, chiral symmetry breaking and confinement occur at the same energy scale. In particular, when $T<T_c$ chiral symmetry is broken and the theory confines, while at $T>T_c$ the theory enters a deconfined phase with chiral symmetry restoration.
However, for non-antipodal configurations with $L M_{KK}<0.966$, an intermediate phase with deconfinement but broken chiral symmetry arises \cite{Aharony:2006da}.
In  the second part of this paper, we will examine this kind of separate first-order phase transition.
The bubble nucleation dynamics is described by the Dirac-Born-Infeld  (DBI) action for the D8-branes on the fixed black brane background. 
 In this case,  even the numerical analysis is challenging due to
 the non-linearities inherent to the DBI action. We will develop a variational approach 
 (which could be hopefully useful to study further static and dynamical issues in the model) to solve the problem. 
This will allow us to compute the (approximate) bounce solution interpolating between the chiral symmetry broken and restored configurations, corresponding to connected and disconnected brane embeddings, and ultimately the actions and decay rates.

Although studying holographic vacuum decay is compelling {\it per se}, it can find an interesting application in connection with gravitational waves, the context in which the analyses related to RS scenarios are typically conducted  \cite{Creminelli:2001th,Randall:2006py,Nardini:2007me,Konstandin:2010cd,Baratella:2018pxi,Megias:2018sxv,Agashe:2019lhy,DelleRose:2019pgi,vonHarling:2019gme,Megias:2020vek}.
First-order phase transitions are quite common in nature and arise in many beyond the Standard Model (BSM) scenarios for the early Universe. First-order cosmological phase transitions can in fact be combined with dynamical mechanisms explaining, for instance, the baryon-antibaryon matter asymmetry or the nature of dark matter. The occurrence of first-order phase transitions in the early Universe would trigger the production of a stochastic gravitational wave background (see e.g. \cite{Caprini:2015zlo, Maggiore:2018sht, Caprini:2019egz}). Provided the transition is strong enough (i.e. provided a relatively large amount of energy is released after the transition), it could  possibly be detected by ground-based as well as space-based future experiments, opening a unique window into BSM physics.

The paper is organized as follows.
In section \ref{secinterlude}, we revisit the compact RS model at finite temperature examined in \cite{Creminelli:2001th}.  Making use of standard holographic renormalization techniques, we compute the kinetic term in the single scalar effective action for the bounce in the deconfined phase. This kinetic term was missing in the literature. 
After devoting section \ref{secWSSreview} to a review of the main features of the WSS model, in section \ref{secWSS} we present the derivation of the effective action for the scalar field modeling the confinement/deconfinement phase transition.
Using holography, we compute both the potential and the kinetic term for the scalar. As in the compact RS example, holographic renormalization techniques play a crucial role in the process. We compute the bubble nucleation rate both in the small temperature regime, where quantum tunneling is driven by $O(4)$-symmetric bubbles, and in the high temperature regime, where $O(3)$-symmetric bubbles are relevant.
In section \ref{sec:chisb} we study the chiral symmetry breaking/restoration phase transition in the deconfined phase. Using a powerful variational method we compute the bounce action and the related bubble nucleation rates. 
In appendix \ref{appthickthin}, we present the thin and thick wall approximations for the confinement/deconfinement phase transition.

In a forthcoming paper \cite{draftph}, we will compute the stochastic gravitational wave spectrum related to cosmological first-order phase transitions having the WSS model as underlying BSM theory.

%%%%%%%%%%%%%%%%%%%%%%%%%%%%%%%%%%%%%%%%%%%%%%%%%
\section{Revisiting the Randall-Sundrum transition}
\label{secinterlude}

In this section, as a warm-up, we revisit the analysis performed in \cite{Creminelli:2001th} of the compact Randall-Sundrum (RS) model with two relevant scales, given by the temperature $T$ and the radial distance between a Standard-Model brane (the TeV brane) and a Planck brane.  
The system experiences a first-order phase transition at some critical temperature $T_c$. At low temperatures, it is described by the RS solution with stabilized radion, while at large temperatures its (bottom-up) holographic description is captured by an AdS$_5$ Schwarzschild black hole whose horizon replaces the TeV brane. A cosmological scenario is considered where the system evolves cooling down from a large temperature stage. The nucleation rate of bubbles of RS vacuum is then estimated. In the process, the horizon radius of the AdS black hole and the radion are promoted to space-dependent fields whose effective action, describing the bounce, is then estimated. Actually, both fields are seen as two different realizations of a single scalar field, whose effective potential can be obtained, in some suitable limit, by gluing the effective potentials in the two phases. In the following section, we will apply the same strategy to model the dynamics of the confinement/deconfinement transition in the top-down WSS model.

Before going on, let us recall that a missing piece in the analysis of \cite{Creminelli:2001th} was the computation of the effective kinetic term for the horizon radius field. Here we present a proposal to fill this gap.
Although in \cite{Creminelli:2001th} the horizon radius field is ultimately not employed, essentially because its contribution is argued to be subleading with respect to the radion, in the subsequent literature on the gravitational wave spectra in this type of models this field is commonly included in the calculations, so the precise normalization of its kinetic term is important (see e.g. \cite{Nardini:2007me, Baratella:2018pxi}).

Let us work in Euclidean signature, with Einstein-Hilbert gravity action given by
\be
S_{EH}= - 2 M^3\int d^5x \sqrt{g}\left[{\cal R} + \frac{12}{L^2}\right]\,,
\label{act5d}
\ee
where $M$ is the 5d Plank mass. The (Euclidean) AdS$_5$ Schwarzschild solution is given by
\be
ds ^2 = \pr{\frac{u}{L}}^2 \pq{f_T(u) dt^2 + d x^ i dx^i} + \pr{\frac{u}{L}}^{-2} \frac{du ^2}{f_T(u)}\,,\quad f_T(u) =1- \frac{u_T ^4}{u^4}\,, \label{adsS}
\ee
where $L$ is the AdS radius. The real-time (Minkowski) metric has an event horizon at $u=u_T$.

In the near-horizon ($u\rightarrow u_T$) limit, the metric of the $(t,u)$-subspace becomes
\be
ds^2 _{(t,u)} = \frac{4 u_T}{L^2} (u-u_T) dt^2 + \frac{L^2}{4 u_T} \frac{du ^2}{u-u_T} \ . 
\ee
By performing the change of variables
\be
r(u)= \frac{L}{\sqrt{u_T}} \sqrt{u-u_T}\ ,\q \quad  \th (t)=2 \pi T t\,, 
\label{changeov}
\ee
we see that the metric is that of a cone
\be
ds^2 _{(t,u)} = (\sin \a)^2 r^2 d \th ^2 + dr^2 \ ,
\ee
with
\be
\sin \a = T_h/T\,,\quad \q T_h\equiv \frac{u_T}{\pi L^2}\ .
\label{conicalAdS}
\ee
When $T_h=T$, there is no conical singularity and the metric is a proper solution of $S_{EH}$. In this case, the free energy density of the black hole is given by
 \be
f_{BH}=-2\pi^4 (M L)^3 T^4\,.
\label{fbhads}
\ee
The above result can be obtained in at least three equivalent ways. The fastest one consists in integrating the thermodynamic relation $ s = - \partial_{T} f$, where $s$ is the Bekenstein-Hawking entropy density. Alternatively, one can use the holographic relation  
\be
F \equiv f\, V_3 = S_{ren}T\,,
\ee
where $F$ is the free energy, $V_3=\int d^3 x$ is the infinite flat 3d space volume and $S_{ren}$ is the renormalized on-shell Euclidean action. The latter, as reviewed in \cite{Creminelli:2001th}, can be obtained as the difference between the on-shell value of the action (\ref{act5d}) on the black hole solution (\ref{adsS}) and its on-shell value on a pure AdS spacetime with the same boundary. 
Alternatively, it can be obtained by 
the procedure of 
holographic renormalization (see e.g. \cite{Skenderis:2002wp} for a review). In the present setup, it amounts to writing
\be
S_{ren} =\lim_{u_{\Lambda}\to\infty} \left[S_{EH} + S_{GH} + S_{ct}\right] = \lim_{u_{\Lambda}\to\infty} \left[S_{EH} + 2 M^3 \int_{u=u_{\Lambda}} d^4 x \sqrt{h}\left(- 2 K + \frac{6}{L}\right) \right]\,.
\ee
Here $u_{\Lambda}$ is a radial cut-off introduced to regularize the on-shell actions and $h$ is the determinant of the metric at the boundary $u=u_{\Lambda}$. The first piece in round parenthesis is due to the Gibbons-Hawking term $S_{GH}$, $K$ being the trace of the extrinsic curvature of the boundary. The second piece is due to the counterterm action $S_{ct}$ which precisely cancels the divergent terms (in powers of $u_{\Lambda}$) from the on-shell value of $S_{EH}+S_{GH}$. As a result, $S_{ren}$ turns out to be finite. Let us recall that a generic counterterm is required to be covariant with respect to the boundary metric.  

According to the holographic correspondence, eq.~(\ref{fbhads}) can be seen as the free energy density of a dual strongly coupled (3+1)-dimensional conformal field theory (CFT), at finite temperature $T=T_h$, in the planar limit. In top-down holography, an infinite class of explicit examples of such CFT arises by considering the low energy dynamics of $N$ D3-branes at the tip of a six-dimensional (Calabi-Yau) cone. The dual description is provided by AdS$_5\times X_5$ backgrounds where $X_5$ is the base of the cone. The master example is provided by $X_5=S^5$, in which case the six-dimensional transverse space is flat and the dual CFT is ${\cal N}=4$ $SU(N)$ Yang-Mills. 
For all such CFT, 
\be
(M L)^3 =  \frac{N^2}{16\pi^2}\,p\ , \quad p = \frac{\pi^3}{V(X_5)} \ ,
\label{holorel}
\ee
where $V(X_5)$ is the volume of $X_5$. In the ${\cal N}=4$ SYM case, $p=1$.

When $T_h\neq T$, the conical singularity contributes to the free energy. It is useful to consider this possibility since, as it will be clear in a moment, it can provide a natural ``off-shell'' description for the background along the phase transition. As described in \cite{Fursaev:1995ef}, it is possible to regularize the singularity with a two-dimensional spherical cap of radius $r\ra 0$, such that its Ricci scalar $\mc{R}_{S^2}$ is $2/r^2$ and its area\footnote{The sphere is glued to the cone in a way such that their tangent vectors match. As a result the area of the spherical cap reads $2 \pi r^2 \int _{\pi/2 + \a} ^{\pi} d \th \sin \th = 2 \pi r^2(1- \sin \a)$ where $\sin\alpha$ is given in (\ref{conicalAdS}).} is $2 \pi r^2 (1-T_h /T)$.  As a result, the contribution of the spherical cap to the on-shell Euclidean gravity action turns out to be given by
\be
S_{cone} = - 2M^3 \int d^{5} x \sqrt{g}\, \mc{R}_{S^2} = -8\pi M^3 \pr{1- \frac{T_h}{T}} V_3 \frac{u_T ^3}{L^3} \ .
\label{sconads}
\ee
Correspondingly, the contribution to the free energy density is given by
\be
f_{cone} = 8\pi^4 (M L)^3 T_h ^4 \pr{1- \frac{T}{T_h}} \, .
\ee
As a result, the total free energy density reads
\be
f = f_{BH} + f_{cone} =  2\pi^4 (M L)^3 \pr{ 3 T_h ^4 - 4 T T_h ^3} \,,
\label{ftotads}
\ee
which is the result obtained in \cite{Creminelli:2001th}. A crucial idea in that paper was to model the dynamics of the first-order phase transition by means of an effective action for a single scalar field. In the deconfined phase, the latter is realized by promoting the parameter $T_h$ to a space-dependent field. This is the reason why we need to develop an ``off-shell'' formalism where we allow $T_h$ to vary taking general values different from $T$. Within this scheme, eq. (\ref{ftotads}) provides the effective potential for the scalar field $T_h$. Consistently, the potential has a minimum in the homogeneous equilibrium configuration with $T_h=T$.

To proceed further, let us first rewrite the AdS-BH metric (\ref{adsS}) in terms of the radial coordinate $r$ defined in (\ref{changeov}) without restricting the change of variables between $u$ and $r$ to the near horizon limit. As a result
\bea
ds^2 &=& \frac{u_T^2}{L^2}\left(1+\frac{r^2}{L^2}\right)^2\left[f_T(r) dt^2 + dx_i dx_i\right] + \frac{4}{L^2}\left(1+\frac{r^2}{L^2}\right)^{-2} \frac{r^2 dr^2}{f_T(r)}\ ,\nonumber \\
f_T(r) &=& 1-\frac{L^8}{(L^2+r^2)^4}\ ,
 \eea
 with $r$ ranging from zero (at the horizon) to infinity. In this coordinate system, the constant $u_T$ factorizes in a very simple way.
 Let us now consider a simple $O(3)$ symmetric deformation of this metric, allowing just $u_T$ to become a function of the 3d radial variable $\rho=\sqrt{x_i x_i}$,
\be
ds^2 = \frac{u_T(\rho)^2}{L^2}\left(1+\frac{r^2}{L^2}\right)^2\left[f_T(r) dt^2 +  d\rho^2 +\rho^2 d\Omega_2^2 \right] + \frac{4}{L^2}\left(1+\frac{r^2}{L^2}\right)^{-2} \frac{r^2 dr^2}{f_T(r)}\,.
\label{defrho}
\ee 
In this way, the metric exhibits a conical singularity for every value of $\r$ whenever $T_h \neq T$. 
In order to compute the effective 4d action for the field $u_T(\rho)$, one can evaluate the total gravity action (including the contribution (\ref{sconads}) from the conical singularity) on the background (\ref{defrho}) and then integrate over the 5d radial variable $r$.\footnote{This way of proceeding is analogous to what is done to obtain the effective action for the radion, see e.g. \cite{Goldberger:1999un}. Here we are just turning off any fluctuation corresponding to the 4d graviton, according to the semiclassical approximation of \cite{Creminelli:2001th} where the bounce is modeled by a single scalar field action.} The deformation gives rise to terms which depend on the derivatives of $u_T(\r)$. The terms that do not depend on these derivatives are not affected by the deformation, since the latter amounts to a coordinate transformation for them. As a result, the expression (\ref{ftotads}), which gives the effective potential for the field $T_h(\rho)= u_T(\rho) / \pi L^2$, is unchanged. 

The kinetic term in the effective action arises from the on-shell value of
\be
S_{kin} =  -2 M^3  \int d^{5} x \sqrt{g}\,\mc{R} \, .
\ee
Actually, this gives rise to contributions proportional to $(\p_\r u_T)^2$ which diverge as $r\rightarrow\infty$. 
Implementing the holographic renormalization procedure, these divergences can be removed by regularizing the above action term with a cut-off $r_{\Lambda}$, adding the counterterm
\be
S_{kin \, ct} = - 2 M^3 \left(-\frac{L}{2 }\right) \int_{r=r_{\Lambda}} d^4x\,   \sqrt{h}\, \mc{R}_h \ ,
\ee
and taking the $r_{\Lambda}\rightarrow\infty$ limit. In the above expression, $h_{mn}$ is the boundary metric at $r=r_{\Lambda}$ and $\mc{R}_h$ is the corresponding Ricci scalar. 

The renormalized kinetic term is thus given by
\be
S_{kin\, ren} = 6 M^3 \frac{4 \pi}{T L} \int d \r \r ^2  (\p_\r u_T ) ^2 \ .
\label{skinads}
\ee
Rewriting the above result in terms of the field $T_h(\rho)$ and taking into account the potential term from (\ref{ftotads}),  we get the total effective Euclidean action
\be
S_{eff}\equiv \frac{S_3}{T} = \frac{4\pi}{T} (M L)^3\int d\rho\,\rho^2\left[6\pi^2 (\partial_{\rho}T_h)^2 + 2\pi^4(3T_h^4 - 4T_h^3 T)\right]\,.
\ee
Using the holographic relation (\ref{holorel}) and formally reintroducing a covariant 4d notation, the latter expression can be rewritten as
\be
S_{eff} =   \frac{N^2}{16\pi^2}\,p \int d^4 x\left[ 6\pi^2 (\partial_{\mu} T_h)^2 + 2\pi^4 (3 T_h^4 - 4 T_h^3 T)\right]\,.
\ee
This formula is the main result of this section:
our analysis determines the relative coefficient between the kinetic and the potential term in the effective action for the ``temperature field'' $T_h(x)$, for the entire class of  strongly coupled planar (3+1)-dimensional CFT with an AdS$_5$ black hole holographic dual.\footnote{Comparing with the notations of e.g.~\cite{Baratella:2018pxi}, we see that our analysis allows to determine their kinetic term coefficient as $c_3= 48 c_2 = 6\pi^2 p$.}  
%%%%%%%%%%%%%%%%%%%%%%%%%%%%%%%%%%%%%%%%%%%%%%%%%%%%%%%%%%%%%
%%%%%%%%%%%%%%%%%%%%%%%%%%%%%%%%%%%%%%%%%%%%%%
\section{The Witten-Sakai-Sugimoto model}
\label{secWSSreview}

The WSS model is a non-supersymmetric (3+1)-dimensional Yang-Mills theory with gauge group $SU(N)$, coupled to $N_f$ fundamental flavors and a tower of  Kaluza-Klein (KK) matter fields \cite{Witten:1998zw,Sakai:2004cn}.\footnote{See \cite{Rebhan:2014rxa} for a concise review.} Our focus will be on the 't Hooft limit of the model, where $N\gg1$, $N_f/N\ll1$ and the 't Hooft coupling $\lambda$ at the KK mass scale $M_{KK}$ is taken to be very large,
$\lambda \gg 1$. 
  The dimensionful parameter $M_{KK}$ also gives the typical mass scale of the glueballs.  The confining string tension $T_s$ is parametrically larger than $M_{KK}^2$ since $T_s \sim \lambda \, M_{KK}^2$.
The non-perturbative IR dynamics of the model shares many relevant features with real-world QCD, including confinement, mass gap and chiral symmetry breaking. Moreover, the WSS theory exhibits a very interesting phase diagram, with a first-order confinement/deconfinement transition and a first-order chiral-symmetry-restoring transition which can happen at different critical temperatures depending on the parameters of the model. Most importantly, in the above-mentioned regime, all these features can be analytically captured by means of a dual classical gravity description with a very precise embedding in string theory.

In the WSS model, all the fields transforming in the adjoint representation of the gauge group arise from the low energy dynamics of $N$ D4-branes wrapped on a circle $S_{x_4}^1$ with coordinate $x_4 \sim x_4 + 2 \pi / M_{KK}$. When we consider the model at finite temperature $T$, the Euclidean time direction is compactified too, $t \sim t + \b = t \sim t + 1/T$, and therefore we have another circle $S_{t}^1$. Each of the $N_f$ fundamental flavor fields is introduced by means of a pair of D8/anti-D8-branes, transverse to $S_{x_4}^1$, separated by a certain distance $L\leq \pi M_{KK}^{-1}$ along that circle. In the original version of the model, there are $N_f$ D8-branes and $N_f$ anti-D8-branes put at antipodal points on $S^{1}$, i.e. such that $L M_{KK} =\pi$. When the flavors are massless, the gauge symmetry on these branes realizes the classical $U(N_f)_L\times U(N_f)_R$ global chiral symmetry of the theory. In the following we will also consider a more general setup where part (if not all) of the flavor branes are not antipodal. In general, there can be several distinct flavor brane pairs as it happens in the recently considered Holographic QCD axion scenario \cite{Bigazzi:2019eks,Bigazzi:2019hav}.

The WSS model has a very well known holographic dual description. When $N_f=0$ the latter is provided by the so-called Witten-Yang-Mills (WYM) solution \cite{Witten:1998zw} which describes the near horizon limit of the background sourced by the $N$ D4-branes. It is a classical solution of  the Type IIA 10d gravity action with a curved metric, a dilaton and a four-form Ramond-Ramond (RR) field strength turned on. At finite temperature, there are actually two competing solutions, related by the exchange of the two $S^1$ circles mentioned above. By computing the free energy, it turns out that at any given temperature $T$ only one of the two backgrounds is energetically favored. Dialing the temperature, the system exhibits a first-order phase transition. 

One of these backgrounds is the black hole one. Considering the case with Euclidean signature, it reads, in string frame:
\bea
\label{wittenbackground}
ds^2 &=& \pr{\frac{u}{R}}^{3/2} \pq{f_T(u) dt^2 + dx^i dx^i +  d x_4 ^2 } +\pr{\frac{R}{u}}^{3/2} \pq{\frac{du^2}{f_T(u)} +u^2 
d \O _4 ^2}\ ,\nonumber \\
f_T(u) &=&1- \frac{u_{T} ^3}{u^3}\ , \quad e^\ff=g_s \pr{\frac{u}{R}}^{3/4}\ ,\quad F_4= \frac{3 R^3 }{g_s} \o_4\ , \quad R^3=\pi g_s N l_s ^3\ .
\eea
The parameter $u_T$ is related to the Hawking temperature $T_h$ by
\be
u_T = \frac{16 \pi ^2}{9} R^3 T_h ^2 \ .
\label{utth}
\ee
The second background is called {\it solitonic} and reads
\bea
\label{solitonicW}
ds^2 &=& \pr{\frac{u}{R}}^{3/2} \pq{dt^2 + dx^i dx^i + f(u) d x_4 ^2 } +\pr{\frac{R}{u}}^{3/2} \pq{\frac{du^2}{f(u)} +u^2 
d \O _4 ^2}\ ,\nonumber \\
f(u) &=& 1 -\frac{u_0 ^3}{u^3}  \ , \q \q \q u_{0} \equiv \frac{4}{9} R^3 M_{h}^2 \ .
\eea
The dilaton and $F_4$ fields keep precisely the same form as in the previous case. 

As we will see in a moment, the two backgrounds are regular, proper solutions of the type IIA gravity action only when $T_h=T$ and $M_h=M_{KK}$. The map between string parameters and field theory ones is given by
\be
g_s l_s = \frac{1}{4 \pi} \frac{\l}{M_{KK }N } \ , \q \q \q  \frac{R^3}{l_s ^2} = \frac{1}{4} \frac{\l}{M_{KK}}\ , 
\label{holomaps}
\ee
where $\lambda$ is the 't Hooft coupling mentioned at the beginning of this section.

The two backgrounds are simply related by $(t,2 \pi T_h) \lra (x_4,M_h)$. Without imposing further constraints, they both exhibit a conical singularity. Indeed, let us consider the $(t,u)$ subspace of the black hole background and let us expand it in the neighborhood of $u=u_T$,
\be
ds^2 _{(u,t)} = \frac{3 u_{T} ^{1/2}}{R^{3/2}} (u-u_{T}) d t ^2  +\frac{R ^{3/2}}{3 u_{T} ^{1/2}} \frac{du^2}{u-u_{T}}  \ .
\ee
By performing the change of coordinates $(t,u) \ra (\th,r)$
given by
\be
\label{changecoordinates}
r(u) =\frac{2}{\sqrt{3}} \pr{ \frac{R^3}{u_T} }^{1/4} \sqrt{u-u_T}\ , \q \quad \th (t) = 2 \pi T t \ , 
\ee
we find
\be
ds^2 _{(u,t)} = \frac{9 u_{T}}{16 \pi ^2 R^{3} T ^2} r^2 d \th ^2  + dr^2 = \pr{\frac{T_h}{T}}^2 r^2 d \th ^2 + dr^2  \ .
\ee
This is the metric of a cone with angle $\a$ given by $\sin \a = T_h/T $.  Analogously, expanding the metric of the solitonic background around $u=u_0$ we find 
\be
ds^2 _{(u,x_4)}  = y^2 ( \sin \b) ^2 d \th ^2 + dy^2  \ ,
\ee
with $\sin \b = M_h/M_{KK} $. As anticipated above, the conical singularity disappears when $T_h=T$ for the first background and $M_h = M_{KK}$ for the second one. For the purposes of this work, and in analogy with the discussion of section \ref{secinterlude}, it will be useful to consider a general ``off-shell" setup in which the backgrounds display the conical singularity.

The solitonic background is dual to the confining phase of the dual gauge theory. The black hole one is instead dual to the deconfined phase.\footnote{It has been argued in \cite{Mandal:2011ws} that this phase is actually not in the same universality class as that of finite temperature Yang-Mills since some discrete symmetries do not match.} As we will review in section \ref{secWSS}, there is a first-order phase transition between the two phases, with a critical temperature $T_c=M_{KK}/2\pi$. When $T<T_c$ (resp.~$T>T_c$) the theory is in a confined (resp.~deconfined) phase. 

Let us now recall what happens when fundamental flavors are added to the model \cite{Sakai:2004cn}. In the 't Hooft limit, the backreaction of the D8/anti-D8 branes on the above-mentioned backgrounds can be neglected and they can thus be treated as probes. One is just left with solving the Euler-Lagrange equations for the D8-brane embedding described by a function $x_4=x_4(u)$ on both backgrounds. 

In the confined phase the solution is such that each D8 and anti-D8 brane pair is actually joined into a single U-shaped configuration. This geometrically realizes the chiral symmetry breaking of the dual field theory. When the branes are taken to be antipodal on the $S_{x_4}^1$ circle, the bottom of the configuration coincides with the bottom of the space $u=u_0$. This means that chiral symmetry breaking and confinement happen at the same energy scale. However, when the branes are not antipodal, they end up joining at some $u_J>u_0$, in which case the two scales are separated. In the standard QCD-like setup with $N_f$ coincident D8-branes and $N_f$ antipodal anti-D8-branes, the model precisely realizes the breaking of $U(N_f)\times U(N_f)$ to the diagonal $U(N_f)$, and the effective action on the D8-branes turns out to reproduce, at low energy, the chiral Lagrangian (with pion decay constant $f_{\pi}\sim \sqrt{N} M_{KK}$) including the Skyrme term. The $\eta'$-like particle, in the model, gets a mass due to the axial anomaly, precisely as expected in QCD.  Quark mass terms can also be turned on. In \cite{Bigazzi:2019eks} a variant of this setup has been considered, by adding a further non-antipodal D8-brane pair (with $L M_{KK}\ll \pi$) corresponding to an extra massless flavor. The related axial symmetry was identified with the $U(1)$ Peccei-Quinn symmetry and the $\eta'$-like particle arising from its breaking was interpreted as a QCD-like axion, 
see also \cite{Bigazzi:2019hav}. 

In the deconfined phase, there are two possible D8-brane embeddings depending on the distance $L$ along the $S^{1}_{x_4}$ circle \cite{Aharony:2006da}. In particular, for fixed physical parameters $M_{KK}$, $L$, we have the following phases depending on the temperature $T$:

\begin{itemize}
\item  If $T<\frac{M_{KK}}{2\pi}$, the theory is confining and chiral symmetry is broken;
\item If $T>\frac{M_{KK}}{2\pi}$, the theory is deconfined and:
\begin{itemize}
\item If $T<\frac{0.1538}{L}$, chiral symmetry is broken;
\item If $T>\frac{0.1538}{L}$, chiral symmetry is preserved.
\end{itemize}
\end{itemize}
Thus, the intermediate phase with deconfinement and chiral symmetry breaking that will be of interest in section \ref{sec:chisb} exists for
\begin{equation}
\frac{M_{KK}}{2\pi}<T<\frac{0.1538}{L} \ .
\label{eq_mkk}
\end{equation}
According to eq.~(\ref{eq_mkk}), the intermediate phase exists if $M_{KK}L< 0.966$. 

\section{Confinement/deconfinement phase transition}
\label{secWSS}
In this section, we study bubble nucleation in the confinement/deconfinement phase transition in the WSS model. We consider a scenario where the WSS theory starts at high temperature and then cools down. Due to the first-order phase transition, bubbles of confining (solitonic) vacuum will start to nucleate within the deconfined (black hole) vacuum. 

\subsection{Free energies of the Witten backgrounds}\label{freeW}
In order to decide which one of the two possible background solutions (\ref{wittenbackground}) and (\ref{solitonicW}) is energetically favored, one has to compute the related on-shell gravity action. This in turn amounts to computing the free energy of the dual field theory, as we have reviewed in section \ref{secinterlude}. As usual, the on-shell gravity action will be holographically renormalized. Let us review some detail of the computation following \cite{Mateos:2007vn,Bigazzi:2014qsa}. The Euclidean renormalized on-shell gravity action is given by
\be
S_{ren} = S_{IIA} + S_{GH}+ S_{ct}\,,
\ee
where
\be
S_{IIA} = - \frac{1}{2 \kappa _{10} ^2}  \int d^{10} x \sqrt{g} \pq{ e^{-2 \ff} \pr{\mc{R} + 4 \p_M \ff \p ^M \ff }
- \frac{1}{2} |F_{4}|^2 } \ ,
\ee
is the relevant truncation of the type IIA gravity action,
\be
S_{GH} = -  \frac{1}{\kappa _{10} ^2}  \int d^9 x \sqrt{h} e^{-2\ff} K\,,
\ee
is the Gibbons-Hawking term and
\be
S_{ct} = \frac{1}{\kappa _{10} ^2} \frac{g_s ^{1/3}}{R} \int d^9 x \sqrt{h} \frac{5}{2} e^{-7 \ff/3}\,,
\ee
is the counterterm action. In the above expressions, $2 \kappa _{10} ^2 = (2 \pi)^7 l_s ^8$, 
$K$ is the extrinsic curvature of a cut-off surface $u=u_\L$,
\be
K = \frac{1}{\sqrt{g}} \p_u \pr{ \frac{\sqrt{g}}{\sqrt{g_{uu}}}}\Big|_{u=u_\L}\,,
\ee
and $h$ is the determinant of the boundary metric at $u=u_\L$.
Summing up all the contributions and taking the $u_\L\rightarrow\infty$ limit, the renormalized on-shell action on the black hole background (\ref{wittenbackground}) turns out to be given by
\be
S_{ren} =- \frac{ \pi V_{S^4} V_4 }{2 \kappa _{10} ^2 g_s ^2 M_{KK}} u_T ^3 \ .
\ee
Here $V_{S^4}$ and $V_4$ are the volumes of the four sphere and the flat four-dimensional space.
The free energy density of the dual theory is therefore 
\be
f_{BH} = - \frac{1}{2} \pr{\frac{2}{3}}^7 \pi ^4 \l N^2 \frac{T_h ^6}{M_{KK} ^2}\,,
\ee
where we have also used the relations (\ref{holomaps}). Substituting $T_h \ra M_h / 2 \pi$ we find the free energy of the solitonic background, 
\be
f_{solitonic} = -  \pr{\frac{1}{3}}^7 \frac{1}{\pi ^2} \l N^2 \frac{M_h ^6}{M_{KK} ^2} \ .
\ee
When $T_h=T$ and $M_h=M_{KK}$, the energy difference reads
\be
f_{solitonic} - f_{BH} = \frac{\l N^2}{\pi ^2 M_{KK} ^2} \pr{\frac{1}{3}}^7 \pq{- M_{KK} ^6 + (2 \pi T)^6}\  .
\ee
As a result, for temperatures $T < M_{KK} / 2 \pi$ the solitonic solution is energetically favored, while for temperatures $T > M_{KK} / 2 \pi$ the black hole solution dominates. At $T = T_c= M_{KK}/ 2 \pi$ the system features a first-order phase transition.

As shown in the previous section, if $T_h \neq T$ and $M_h \neq M_{KK}$, the backgrounds display a conical singularity and the latter contributes to the free energy. For the black hole background, we regularize the $(t,u)$ subspace smoothing it with a two-dimensional spherical cap precisely as done in the RS-AdS case revisited in section \ref{secinterlude}.
The contribution of the spherical cap to the action is therefore
\be
S^{cone} _{BH} = - \frac{1}{2 \kappa _{10} ^2} \int d^{10} x \sqrt{g} e^{-2 \ff} \mc{R}_{S^2} = -  \frac{2 \pi V_3 V_{S^4} }{2 \kappa _{10} ^2 g_s ^2 M_{KK}} 4 \pi \pr{1- \frac{T_h}{T}} \pr{\frac{u_T}{R}}^{-3/2} u_T^4 \ .
\ee
Analogously, for the solitonic background we have
\be
S^{cone} _{solitonic} = - \frac{1}{2 \kappa _{10} ^2} \int d^{10} x \sqrt{g} e^{-2 \ff} \mc{R} = - \frac{1}{2 \kappa _{10} ^2} \frac{V_3 V_{S^4} \b }{g_s ^2} 4 \pi \pr{1- \frac{M_h}{M_{KK}}} \pr{\frac{u_0}{R}}^{-3/2} u_0^4 \ .
\ee
The contribution of the conical singularity then reads
\ba
f^{cone} _{BH} &=& 3 \pr{\frac{2}{3}}^7 \pi ^4 \l N^2 \frac{T_h ^6}{M_{KK} ^2} \pr{1- \frac{T}{T_h}} \ ,  \\
f^{cone} _{solitonic} &=& 6 \pr{\frac{1}{3}}^7 \frac{1}{\pi ^2} \l N^2 \frac{M_h ^6}{M_{KK} ^2} \pr{1- \frac{M_{KK}}{M_h}} \ .
\ea
As a result, the total free energies read
\ba
f_{BH} ' &=& f_{BH} + f^{cone} _{BH} = \frac{1}{2} \pr{\frac{2}{3}}^7 \pi ^4 \l N^2 \frac{1}{M_{KK} ^2} \pr{5 T_h ^6 - 6 T T_h ^5} \ , \label{poth}\\
f_{solitonic} '  &=& f_{solitonic} + f^{cone} _{solitonic} =  \pr{\frac{1}{3}}^7 \frac{1}{\pi ^2} \l N^2 \frac{1}{M_{KK} ^2} \pr{5 M_h ^6 - 6 M_{KK} M_h ^5} \ .
\ea

\subsection{Holographic bubbles}
In order to describe the nucleation of bubbles, we should find a solution of the equations of motion that interpolates between the confined and the deconfined backgrounds. Unfortunately, this is a very difficult task to pursue. The idea is then to take an effective approach in which the interpolation is mediated by a single effective degree of freedom \cite{Creminelli:2001th}. Since the two backgrounds differ only for the fact that the blackening factor sits in front of $dx_4 ^2$ or $dt^2$, we might try to promote the parameters $u_T$ and $u_0$ to fields $u_T(\r)$ and $u_0(\r)$, where $\r$ is the radial coordinate for the bubble. We will consider either $O(3)$ symmetric bubbles, for which $\rho^2=x_i x_i$, or $O(4)$ symmetric ones, where $\rho^2=t^2+x_i x_i$. For instance, in the black hole case, one could start from a $O(3)$-symmetric ansatz of the form
\be
\label{firstansatz}
ds^2 = \pr{\frac{u}{R}}^{3/2} \pq{f_T(u,\r) dt^2 + d \r ^2 + \r ^2 d \O _2 ^2 +  d x_4 ^2 } +\pr{\frac{R}{u}}^{3/2} \pq{\frac{du^2}{f_T(u,\r)} +u^2 
d \O _4 ^2} \ ,
\ee
with
\be
f_T(u,\r) =1- \frac{u_{T} (\r) ^3}{u^3}\,,
\ee
and the other fields left unchanged. 
In this setup, the temperature of the horizon $T_h$ is promoted to a field as well,
\be
u_T(\r) = \frac{16 \pi ^2}{9} R^3\, T_h (\r) ^2  \ .
\label{utthrho}
\ee
The effective action for this field will now include a contribution from its kinetic term. This comes from the Ricci scalar and reads
\ba
\label{riccinonconical}
\mc{R}_{kin} &=& - \frac{9}{2} \pr{\frac{u}{R}}^{3/2} \frac{R^3 u_T ^4}{u^3 (u^3 - u_T ^3)^2} (\p_\r u_T )^2 \nb \\
&=& - \frac{9}{2} \pr{\frac{32 \pi ^2}{9}}^2 \pr{\frac{u}{R}}^{3/2} \frac{R^9 u_T ^4}{u^3 (u^3 - u_T ^3)^2} T_h ^2 (\p_\r T_h)^2 \ .
\ea
Thus we see that using the ansatz (\ref{firstansatz}) the Ricci scalar (\ref{riccinonconical}) displays a divergence for $u \ra u_T(\r)$ which deviates from the conical singularity. Indeed, if we expand the metric around $u=u_T(\r)$, we do not find the metric of a cone, because the change of coordinates (\ref{changecoordinates}) becomes non-trivial when $u_T$ is a function of $\r$. This background is not satisfactory, because we would like it to display a conical singularity with a $\r$-dependent cone angle.

Let us consider another ansatz. We start from the background (\ref{wittenbackground}) and we perform the coordinate change between $u$ and $r$ as in (\ref{changecoordinates}). Then we promote $u_T$ to be a function of $\r$. In this way, the metric expanded around $r=0$ is the metric of a cone for any value of $\r$.  In general, it reads
\be
\label{secondansatz}
ds^2 = \pr{\frac{u}{R}}^{3/2} \pq{f_T(u) dt^2 + d\rho^2 +\rho^2 d\Omega_2^2 +  d x_4 ^2 } +\pr{\frac{R}{u}}^{3/2} \pq{\frac{9\,u_T\, r^2 dr^2}{4 R^3 f_T(u)} +u^2 
d \O _4 ^2} \ ,
\ee
where 
\be
u = u(r,\rho) = u_T(\rho) + \frac{3}{4} \sqrt{\frac{u_T(\rho)}{R^3}} r^2 \,.
\label{redef1}
\ee
The dilaton and the RR four form will be taken as in the original background. In particular, due to eq. (\ref{redef1}), the dilaton will now be a function of both $r$ and $\rho$.

The effective four-dimensional action for $u_T(\rho)$ will be obtained by plugging the ansatz above in the renormalized action $S_{ren}= S_{IIA}+ S_{ct} + S_{GH}$ as defined in section \ref{freeW} and integrating over $r, x_4$ and the transverse four-sphere. The background deformation described above affects only the quantities which depend on the derivatives of $u_T (\r)$, namely the kinetic term of the effective action.
Thus, the potential term in the effective action will be read from eq.~(\ref{poth}) where $T_h(\rho)$ is expressed in terms of $u_T(\rho)$ by means of eq.~(\ref{utthrho}).

The kinetic term in the effective action for $u_T(\rho)$ requires some care. In principle, it is obtained from the on-shell value of
\be
S_{kin\,eff} =  - \frac{1}{2 \kappa _{10} ^2}  \int d^{10} x \sqrt{g} \pq{ e^{-2 \ff} \pr{\mc{R} + 4 \p_{\rho} \ff \p ^{\rho} \ff }}\,.
\ee
Actually, this gives rise to contributions proportional to $(\p_\r u_T  (\rho))^2$ which diverge as $r\rightarrow\infty$. Remarkably enough, the above divergences can be removed by adding to the action above the counterterm
\be
S_{kin \, ct} =  - \frac{1}{2 \kappa _{10} ^2 } \left(-\frac{40 R}{9 g_s ^{1/3}}\right) \int_{r=r_{UV}} d^9x\, \sqrt{h}\, e^{-5\phi/3}\,h^{mn}\,\partial_{m}\phi \,\partial_n \phi\ ,
\ee 
where $h_{mn}$ is the boundary metric at fixed $r=r_{UV}$. All in all we get a quite simple effective action for $u_T(\rho)$. 

It is possible to show that precisely the same results (and the same expression for the renormalized kinetic term) can be obtained using an alternative counterterm action that is built having in mind the structure of the first two terms of the counterterm action in eq.~(5.78) of \cite{ksken}. It reads
\bea
S_{kin\,ct\,alt} &=& -\frac{1}{2\kappa_{10}^2}\left(-\frac{5R}{7 g_s^{1/3}}\right)\int_{r=r_{UV}}
d^9x\, \sqrt{h}\, e^{-5\phi/3}\, {\cal R}_{[h]}+\nonumber \\
&& -\frac{1}{2\kappa_{10}^2}\left(\frac{60}{7 R\, g_s^{-1/3}}\right)\int_{r=r_{UV}}
d^9x\, \sqrt{h}\, e^{-7\phi/3}\,.
\eea
The second, ``volume'' counterterm, cancels all the divergences and the finite terms - which do not depend on derivatives of $u_T(\rho)$ - coming from the first one. The structure of this term is analogous to that of the ``volume'' counterterm we have added to renormalize the bulk on-shell action. 

With the same procedure we can get an effective action for $u_0(\rho)\sim M_h(\rho)^2$ in the confined phase. 

The ansatz we have chosen in our discussion above is $O(3)$ symmetric. This is what is expected to hold at large enough temperatures. For smaller temperatures, one should expect a $O(4)$-symmetric ansatz to hold. This ansatz would be perfectly consistent with the symmetries of the solitonic background dual to the confined phase. In fact, even on the black hole background, which has only $O(3)$ symmetry, at small enough temperature the radius of the bubble can be much smaller than the length of the time circle. In this case, the configuration can effectively enjoy an enlarged $O(4)$ symmetry including the Euclidean time direction \cite{Linde:1980tt,Linde:1981zj}.
We will present the related effective actions in the following subsection.

%%%%%%%%%%%%%%%%%%%%%%%%%%%%%%%%%%%%%%%%%%%%%%%%%%%%%%%%%
\subsection{Effective actions and solutions}
\label{effectiveactionsection}

Let us now write the effective actions for $u_T(\rho)$ or $u_0(\rho)$ in terms of  the field 
\be
Y=-Y_T ({\rm deconfined\, phase}) \ , \quad Y=Y_0 ({\rm confined\, phase}) \ ,
\ee
where
\be
Y_T= T_h(\rho)^2 \ ,\quad Y_0= \pr{\frac{M_h(\rho)}{2 \pi}}^2\ .
\ee
In the $O(3)$-symmetric case, the effective action in the deconfined phase reads
\be\label{o3dold}
\frac{S_3 (Y)}{T} = \frac{16\pi^3 \lambda N^2}{3^5 M_{KK}^2 T}\int d\rho \rho^2 \left[\left(5-\frac{\pi}{2\sqrt{3}} \right) Y'^2 - \frac{16\pi^2}{9}\left(5 Y^3 + 6 T (-Y)^{5/2} \right)   \right] \ ,
\ee
where the prime denotes derivative with respect to $\rho$, and $Y$ is supposed to take negative values. 
In the confined phase the action is
\be\label{o3cold}
\frac{S_3 (Y)}{T} =  \frac{16\pi^3 \lambda N^2}{3^5 M_{KK}^2 T}\int d\rho \rho^2 \left[\left(5-\frac{\pi}{2\sqrt{3}} \right) Y'^2 +\frac{16\pi^2}{9}\left(5 Y^3 - \frac{3}{\pi}M_{KK} Y^{5/2} \right)   \right] \ ,
\ee
where now $Y$ takes positive values.
The full problem is simply the junction of the two regimes. By passing to dimensionless quantities
\be
\label{dimsq}
\Phi \equiv  \frac{Y}{M_{KK}^{2}}\ , \q \quad \bar \rho \equiv M_{KK} \rho\,, \q \quad \bar T \equiv \frac{2\pi T}{M_{KK}}\ , 
\ee
such that the critical temperature $T_c$ corresponds to $\bar T=1$, one factorizes the parametric dependences out of the Lagrangians and the whole action reads
\be
\frac{S_3 (\Phi)}{T}=\frac{32\pi^4 g}{3^5 \bar T} \int_{0}^{\infty}d\bar\rho \bar\rho^2 \left [ \left(5-\frac{\pi}{2\sqrt{3}} \right) \Phi'^2 +  \Theta(\Phi) V_c(\Phi) + \Theta(-\Phi) V_d(\Phi) \right]\,,
\label{o3dc}
\ee
where $\Theta(\cdot)$ is the Heaviside step function, 
\bea
V_c(\Phi) &=& \frac{16\pi^2}{9}\left(5\Phi^3-\frac{3}{\pi}\Phi^{5/2}\right)\,,\nonumber \\
V_d(\Phi) &=& -\frac{16\pi^2}{9}\left(5\Phi^3+\frac{3}{\pi}\bar T (-\Phi)^{5/2}\right)\,,
\eea
and
\be
g \equiv \lambda N^2\ .
\ee
Formula (\ref{o3dc}) is the main result of this section, providing the action for the scalar field effectively describing the interpolation between the black brane and solitonic backgrounds. Note that there is a single parameter $g$ which enters multiplicatively the action. 

\begin{figure}
\center
\includegraphics[scale=1.3]{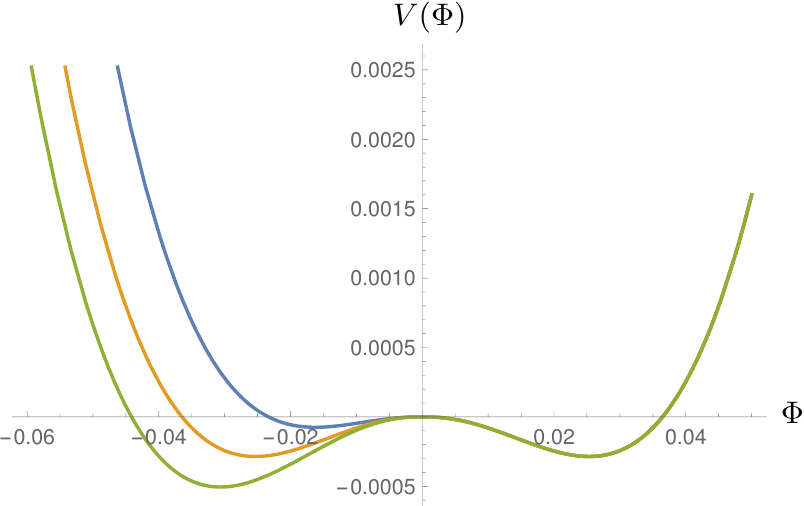}
\caption{Representative curves of the potential for three different values of the dimensionless temperature: $\bar T=0.8$ (blue), $\bar T=1$ (orange), $\bar T=1.1$ (green). The region where $\Phi$ takes positive values does not depend on the temperature, hence the curves overlap.}
\label{figpot}
\end{figure}

Figure \ref{figpot} depicts the full potential for three different values of the reduced temperature $\bar T$. The two minima are $V_d=-\bar T^6/(36 \pi^4)$ for $\Phi_d=-\bar T^2/(4\pi^2)$ and $V_c=-1/(36 \pi^4)$ for $\Phi_c= 1/(4\pi^2)$. We will focus on the case $\bar T \in [0,1]$, where the true vacuum is the confining one at $\Phi=\Phi_c$.

We are going to find a bubble-like solution $\Phi_B$ of the equation of motion derived from the action (\ref{o3dc}) in the following way.
We start inside the bubble, i.e.~for $\bar\rho \in [0,\bar\rho_w]$ (where $\bar\rho_w$ is the location of the bubble wall), i.e.~in the confined case with $\Phi>0$.
The equation is solved with boundary conditions
\be
\Phi _B (0)=\Phi_0 \ , \q \q \q \Phi' _B(0)=0 \ ,
\ee
for some positive value $\Phi_0$; the second condition corresponds to the request of regularity.
The solution $\Phi _B$ is going to vanish at a finite position of the radius, which is identified with $\bar\rho_w$.
There we calculate the derivative $\Phi' _B(\bar\rho_w)\equiv \Phi'_{B, w}$.

Then we solve the equation outside the bubble, i.e.~for $\bar\rho \in [\bar\rho_w, \infty]$, i.e.~in the deconfined case where $\Phi<0$. 
The boundary conditions we use are the ones enforcing continuity of $\Phi _B$ and $\Phi' _B$ at the junction, 
\be
\Phi _B(\bar\rho_w)=0 \ , \q \q \q \Phi' _B(\bar\rho_w)=\Phi'_{B,w} \ .
\ee

Finally, we search for the initial value $\Phi_0$ at the center of the bubble such that the solution for large $\bar\rho$ goes to the false vacuum, $\Phi_d$.
Thus, the whole solution is such that at the center of the ball it goes to a positive constant\footnote{Note that the constant $\Phi _0$ is typically different from the true vacuum $\Phi_c$, because the equation of motion derived from (\ref{o3dc}) contains a friction term.} with vanishing derivative and at infinity it goes to the false vacuum solution. 
\begin{figure}
\center
\includegraphics[scale=1]{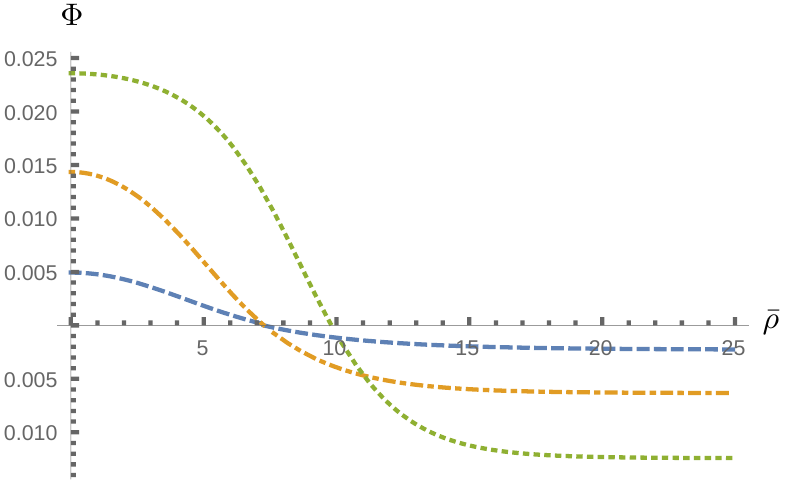} \includegraphics[scale=1]{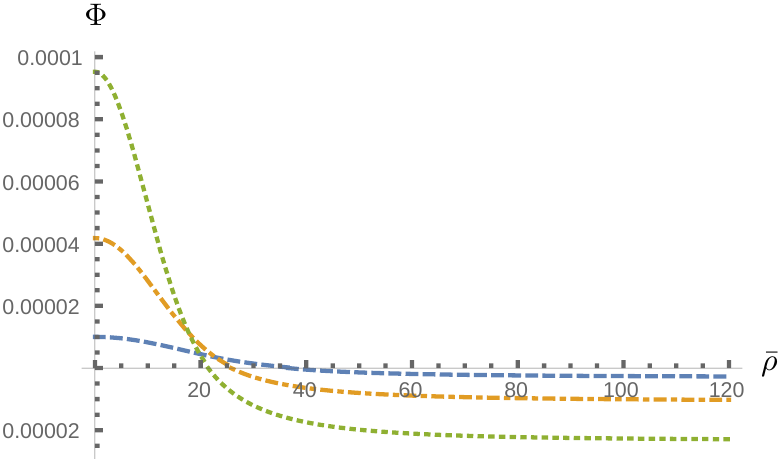}
\caption{Solutions for the bubble profile in the $O(3)$ case (left) with $\bar T=0.3$ (dashed), 0.5 (dash-dotted), $0.7$ (dotted) and in the $O(4)$ case (right) with $\bar T=0.01$ (dashed), 0.02 (dash-dotted), $0.03$ (dotted).}
 \label{figsol}
\end{figure}
Examples of solutions corresponding to different choices of $\bar T$ are given in figure \ref{figsol}.
The amplitude of the configuration is reduced as the temperature gets smaller and smaller. 

Once the solution is calculated, one can plug it back in the action.
As mentioned in the introduction, the bounce action $S_B$ that enters the formula $\Gamma= A\, e^{-S_B }$ for the rate of the vacuum decay is, in the $O(3)$-symmetric case, $S_B = S_{3,B}$ given by \cite{Coleman:1977py}
\be
\frac{S_{3,B}}{T} = \frac{S_3(\Phi_B) - S_3(\Phi_d)}{T}  \ .
\ee

For small temperatures, one could have also $O(4)$ symmetric bounces.
The action is almost the same as (\ref{o3dc}), but for the fact that the four-dimensional measure $d ^4 x$ is now given by $d \O_3 d \r \r ^3$, where $d \O_3$ is the measure of the three-sphere. As a result, the action does not display the overall $M_{KK}/T=2\pi/\bar T$ factor that in the $O(3)$ came from the integration over $t$,
\be\label{o4dc}
S_4(\Phi ) = \frac{8\pi^4 g}{3^5} \int_{0}^{\infty}d\bar\rho\,\bar\rho^3 \left[ \left(5-\frac{\pi}{2\sqrt{3}} \right) \Phi'^2 + \Theta(\Phi) V_c(\Phi) + \Theta(-\Phi) V_d(\Phi) \right]\ . 
\ee
Then, proceeding as above, one obtains solutions for the bubbles as in figure \ref{figsol}.
The bounce action is defined as $S_{4,B} = S_4(\Phi_B) - S_4(\Phi_d)$.

In appendix \ref{appthickthin} we report on the use of the thin and thick wall approximations, which allow us to study semi-analytically the problem at large and small temperatures, respectively.
There it is also shown that the bubble is unlikely to have an even larger symmetry than $O(4)$.
In fact, in principle in the dual description, the bubble could happen to be small as compared to the four-sphere and the $x_4$ circle of the background.
In appendix \ref{appthickthin} we show that this is never the case for temperatures below $T_c$, justifying the ansatze adopted in this section. 

\begin{figure}
\center
\includegraphics[scale=0.98]{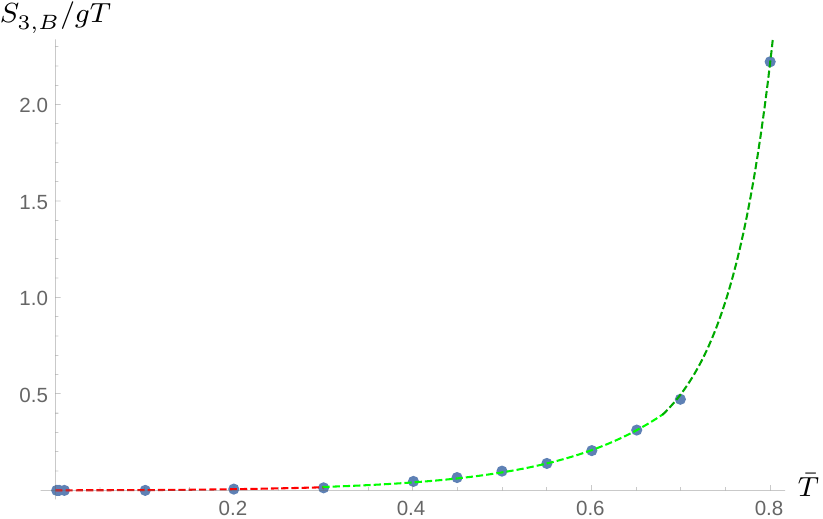} \includegraphics[scale=0.98]{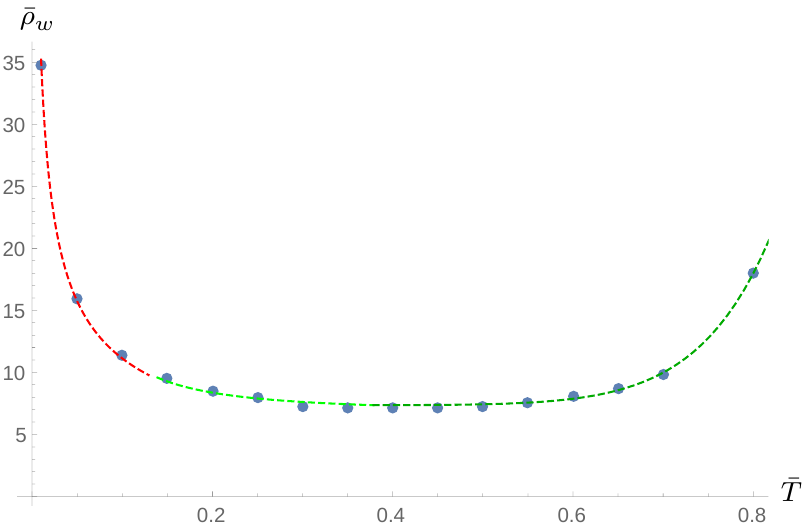}
\caption{The action $S_{3,B}/g T$ and dimensionless radius $\bar\rho_w$ of the $O(3)$ symmetric bubble as a function of $\bar T$. Dots correspond to numerical results, the dotted lines to eqs.~(\ref{fitsO3}), (\ref{fitsRadO3}). Different colors correspond to different expressions of the piecewise functions.}
 \label{figradO3cd}
\end{figure}

Based on the numerical results and 
inspired by the functional form of the thin and thick wall approximations studied in appendix \ref{appthickthin},
a continuous analytic approximation to the action for the $O(3)$ bubble can be provided as follows,
\begin{equation}
\frac{S_{3,B} }{g T} \approx \begin{cases}
0.32\ \bar{T}^{5/2} \qquad\qquad\qquad\qquad\qquad\qquad\ (\bar{T} \leq 0.3)\\
1.8 \times 10^{-3} \exp (7.9\ \bar T) - 2\times 10^{-3} \qquad ( 0.3\leq \bar T  \leq 0.68)\\
5.4 \times 10^{-2} \exp (8.8\ \bar T^{3.8}) \qquad\qquad \qquad   ( 0.68\leq \bar T  \leq 0.87)\\
2.6/\bar T(1\, -\bar T^6)^2\qquad \qquad\qquad\qquad \quad \, \,  \, (\bar T \geq 0.87)
\end{cases}
\label{fitsO3}
\end{equation}
while its radius can be approximated as
\begin{equation}
\bar \rho_w \approx \begin{cases}
3.5 / \bar{T}^{1/2} \qquad \qquad\qquad\qquad\qquad\, \,\,(\bar{T} \leq 0.13)\\
6.8 + 0.13/ \bar{T}^{1.5}  \  \ \qquad\qquad\quad  \qquad ( 0.13\leq \bar T  \leq 0.38)\\
7.4 + 110\ \bar{T}^{10} \qquad\qquad\qquad  \qquad\, ( 0.38\leq \bar T  \leq 0.84)\\
16/(1\, -\bar T^6)\qquad \, \qquad\qquad\qquad \, \, \,  (\bar T \geq 0.84)
\end{cases}
\label{fitsRadO3}
\end{equation}
Figure \ref{figradO3cd} shows a comparison between the latter fits and numerical data.

\begin{figure}
\center
\includegraphics[scale=0.98]{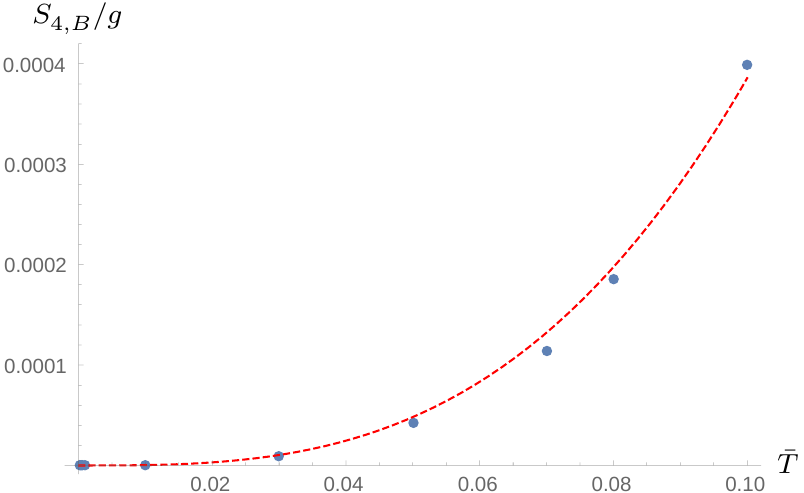} \includegraphics[scale=0.98]{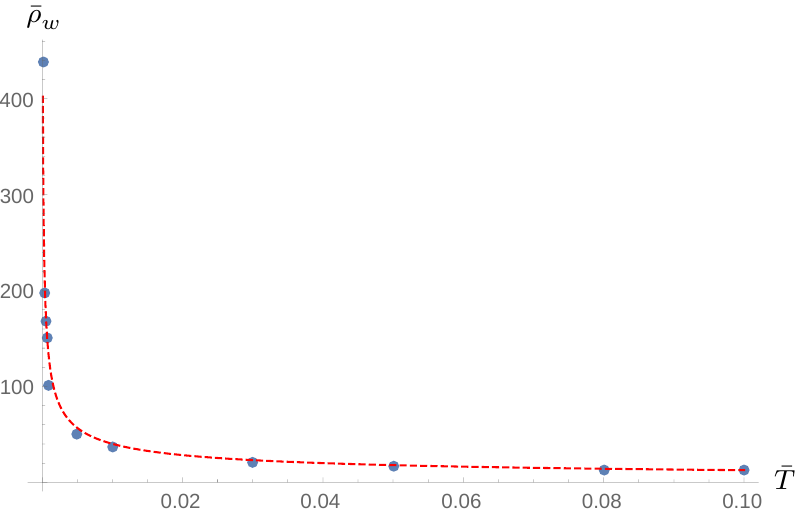}
\caption{The action $S_{4,B}/g$ and dimensionless radius $\bar\rho_w$ of the $O(4)$ symmetric bubble as a function of $\bar T$. Dots correspond to numerical results, the dotted lines to eq.~(\ref{fitsO4}).}
 \label{figradO4cd}
\end{figure}

For the $O(4)$ bubble, since it is only defined for small temperatures, it is sufficient to consider 
the functional form of
the thick wall approximation, giving
\begin{equation}
\frac{S_{4,B} }{g} \approx 0.39\ \bar{T}^{3} \ , \qquad \qquad \bar \rho_w \approx \frac{4.0}{\bar T^{1/2}}  \qquad\qquad\ (\bar{T} < 0.06) \ .
\label{fitsO4}
\end{equation}
The comparison with numerical data is shown in figure \ref{figradO4cd}.
We only plot $S_{4,B}/g$ for small $\bar T$ because of its range of validity.
In fact, the $O(4)$ bubble radius must be much smaller than $1/T$, otherwise one cannot have this enlarged symmetry configuration on the thermal circle \cite{Linde:1980tt,Linde:1981zj}.\footnote{The $O(4)$ bubble does not fit the thermal circle for $2\rho_w > 1/T$. But even if $2\rho_w < 1/T$, if the radius is close to the extremal value $1/2T$, the assumption that there is an enlarged $O(4)$ symmetry is hardly consistent.}
We choose to place the discriminant bubble radius value, above which we do not consider $O(4)$ configurations, at the conventional point where $\rho_w = 1/2\pi T$ (the radius of the thermal circle).
In our case, this happens for $\bar T \approx 0.06$.

%%%%%%%%%%%%%%%%%%%%%%%%%%
\subsection{Bubble nucleation rate}
\begin{figure}
\center
\includegraphics[scale=1]{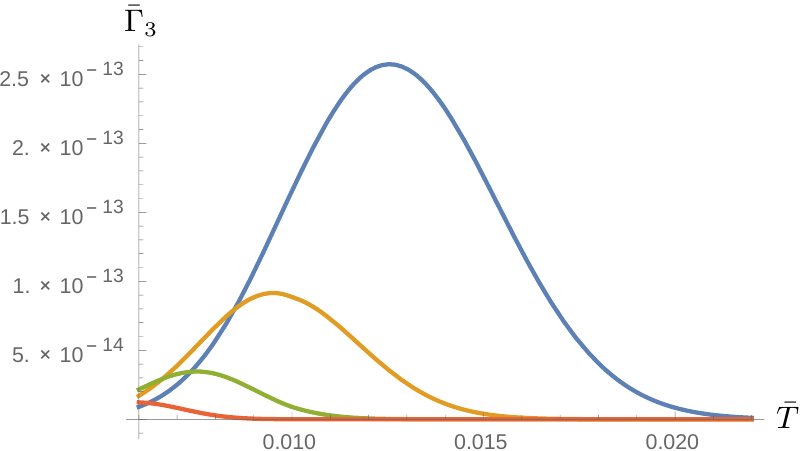}\includegraphics[scale=1]{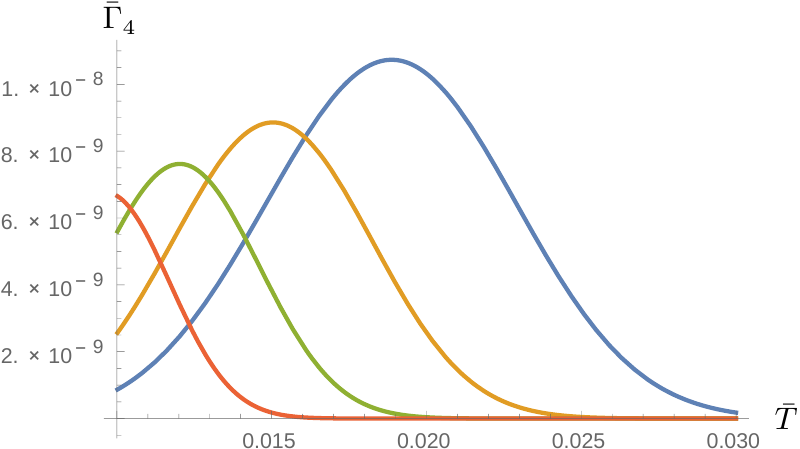}
\caption{Representative plots of the decay rate $\bar\Gamma \equiv \Gamma/M_{KK}^4$ for $g/10^6=$ 1 (blue), 2 (orange), 4 (green), 8 (red), for the $O(3)$ (left) and $O(4)$ (right) configurations.}
 \label{figrates3}
\end{figure}

The bubble nucleation rate is the maximum of the rates of the $O(3)$ and $O(4)$ symmetric bubbles
\cite{Coleman:1977py,Callan:1977pt,Coleman:1980aw,Linde:1980tt,Linde:1981zj}\footnote{The prefactors $T^4$ and $1/\rho_w^4$ in (\ref{Gamma}) are essentially determined by dimensional analysis and heuristic considerations \cite{Callan:1977pt,Linde:1981zj}. We verified that changing e.g.~$T^4$ into $T^6/M_{KK}^2$ has very small impact on the numerical values found in this paper.} 
\bea \label{Gamma}
\Gamma & = & {\rm Max}\left[T^4 \left( \frac{S_{3,B} }{2\pi T} \right)^{3/2} e^{-S_{3,B} /T}      ,  \left( \frac{S_{4,B} }{2\pi \rho_w^2}  \right)^2  e^{-S_{4,B} }   \right]   \nonumber \\
& = & M_{KK}^4 {\rm Max}\left[\frac{\bar T^4}{(2\pi)^4} \left( \frac{S_{3,B}}{2\pi T} \right)^{3/2} e^{-S_{3,B}/T}      ,  \left( \frac{S_{4,B}}{2\pi \bar\rho_w^2}  \right)^2  e^{-S_{4,B}}   \right] \,. 
\eea
Some examples of the rates in the $O(3)$ case are provided in figure \ref{figrates3}.
Since the rate is exponentially suppressed with the action, it is more and more suppressed as the parameter $g$ is increased.
Also, the peak of the rate is shifted to smaller temperatures by increasing $g$, so that for large values of this parameter the theory features what is called {\it supercooling}. In this case, the rate is so small that the theory is trapped in the false vacuum, below the critical temperature of the first-order transition, for a long time.

Similar features are present in the $O(4)$ case, shown again in figure 
\ref{figrates3}.
As can be appreciated by comparing the left and right plots in figure \ref{figrates3}, which correspond to the same values of $g$, the rate for the $O(4)$ bubble dominates on the one for the $O(3)$ bubble for those values of $\bar T$ for which it is defined, namely for $\bar T \lesssim 0.06$.
Thus, at such small temperatures, the decay is much more likely to happen via quantum rather than thermal fluctuations.

%%%%%%%%%%%%%%%%%%%%%%%%%%%%%%%%%%%%%%%%%%%%%%%%%%%%%%%%%%%%%%%%%%%
\section{Chiral symmetry phase transition}
\label{sec:chisb}

\subsection{Revisiting the transition}
\label{RevisitingchiSB}

As already mentioned in section \ref{secWSSreview}, the authors of \cite{Aharony:2006da} showed that in the Witten-Sakai-Sugimoto model
the deconfinement phase transition and the chiral symmetry breaking phase transition can take place at different temperatures for certain parameters of the model.
Thus, apart from the vacuum decay studied in section \ref{secWSS}, there is a different type of vacuum decay associated to the embedding of the flavor branes.
In this section we will briefly review the analysis of \cite{Aharony:2006da}
and then put forward a simple analytic expression that approximates
with good accuracy the brane embedding profiles. This expression will be a useful tool in subsection \ref{subsec:flavor_bubbles} where we will
discuss the bubble configurations that mediate the chiral symmetry breaking phase transitions in the deconfined phase.

We want to study probe brane embedding profiles in the
(Euclidean) background given by Eqs.~(\ref{wittenbackground}) where one must take into account that
\begin{equation}
u_T=\frac{16\pi^2}{9}R^3T^2 \ ,\qquad x_4 \sim x_4 + \frac{2\pi}{M_{KK}} \ , \qquad  t \sim t + \frac{1}{T} \ .
\label{uTval}
\end{equation}
 
\begin{figure}[htb]
\begin{center}
\includegraphics[width=0.49\textwidth]{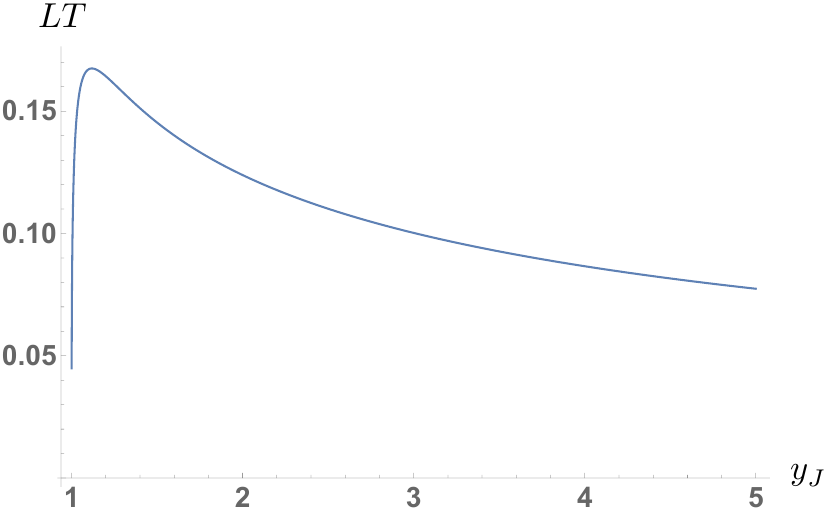}
\includegraphics[width=0.49\textwidth]{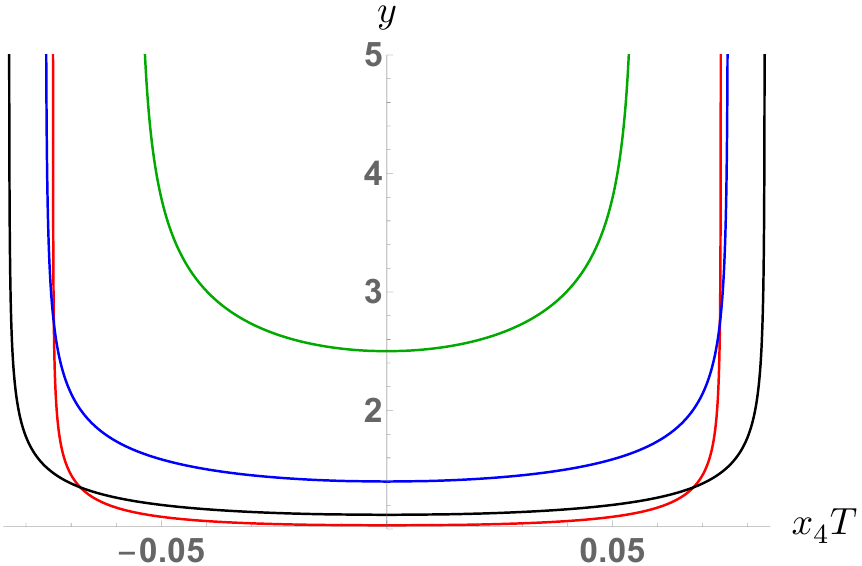}
\end{center}
\caption{Left: Separation in $x_4$ times the temperature as a function of $y_J=u_J/u_T$ for the U-shaped configuration in the deconfined
background. The maximum value of $L\,T$ in the plot is $L\,T \approx 0.1675$ and occurs for $y_J \approx 1.119$.
Right: Profiles for different values of $y_J$: $y_J=1.03$ (red), $y_J=1.119$ (black), $y_J=1.4$ (blue), $y_J=2.5$ (green). We have
assumed, without loss of generality, that the tip of the brane is located at $x_4=0$.}
\label{fig_LT}
\end{figure}

 The Sakai-Sugimoto model \cite{Sakai:2004cn} consists in  introducing D8 probe flavor branes
 extended along the Minkowski directions, the four-sphere and $u$, with a profile $x_4 = x_4(u)$. The Dirac-Born-Infeld action is
\begin{equation}
S_{DBI}=\frac{T_8}{g_s}\int d^9x \left(\frac{u}{R}\right)^{-3/2}u^4 \sqrt{1 + f_T(u) \left(\frac{u}{R}\right)^3 (\partial_u x_4)^2}  \ .
\label{SDBI1}
\end{equation}
From the Euler-Lagrange equation $\partial_u \left(\frac{\partial {\cal L}}{\partial_u x_4} \right)= \frac{\partial {\cal L}}{\partial x_4} $, we find that 
$\frac{\partial {\cal L}}{\partial_u x_4}$ is a constant and therefore
\begin{equation}
 \left(\frac{u}{R}\right)^{-3/2}u^4 \frac{f_T(u) \left(\frac{u}{R}\right)^3 (\partial_u x_4)}{\sqrt{1 + f_T(u) \left(\frac{u}{R}\right)^3 (\partial_u x_4)^2} } = \text{constant} \ .
 \label{simple1}
\end{equation}
The simplest solution is that of a straight brane-antibrane pair each at constant $x_4$. That would be the phase with unbroken chiral symmetry.
On the other hand, there are U-shaped solutions that connect the brane and the antibrane somewhere in the bulk, leading to a breaking
of chiral symmetry.
Any solution of that kind has a tip, where the brane and antibrane are joined, located at some position of the holographic direction $u=u_J$ such that $x_4'(u_J)=\infty$.
For this case, we can rewrite (\ref{simple1}) as
\begin{equation}
\frac{u^4 \sqrt{f_T(u)}}{\sqrt{1 + \left(f_T(u) \left(\frac{u}{R}\right)^3 (\partial_u x_4)^2\right)^{-1}}} = u_J^4 \sqrt{f_T(u_J)} \ .
\label{simple2}
\end{equation}
We can rescale the coordinate to factor out the dimensionful parameters,\footnote{Notice that the $y$ defined here does not coincide with the one defined by \cite{Aharony:2006da}.}
\begin{equation}
x_4 = x \,  u_T^{-1/2} R^{3/2} = x \frac{3}{4\pi T} \ , \qquad u = y\,u_T   \ ,  \qquad u_J = y_J\,u_T\,,
\label{redef}
\end{equation}
such that
\begin{equation}
f_T(u) \equiv f_T = 1 - y^{-3}  \ , \qquad  f_T(u_J)  \equiv f_{TJ} = 1 - y_J^{-3} \ .
\label{redef2}
\end{equation}
The periodicity of the cigar coordinate is
\begin{equation}
x \sim x +  \frac{2\pi\sqrt{ u_T}}{M_{KK} R^\frac32}=
x +  \frac{8\pi^2 T}{3 M_{KK}} \ .
\end{equation}
In these coordinates, equation (\ref{simple2}) can be rewritten as
\begin{equation}
\partial_y x = \left[f_T  y^3 \left(\frac{y^8f_T}{y_J^8 f_{TJ}} - 1 \right)  \right]^{-1/2} \ .
\label{yofx1}
\end{equation}
Recalling that $L$ is  the distance between the brane and the antibrane along $x_4$ in the $u\to \infty$ limit,
for the U-shaped configuration, it can be computed as
\begin{equation}
L = \int_{worldvolume} dx_4 = 2 \int_{u_J}^\infty \frac{dx_4}{du} du = 2 \frac{3}{4\pi T} \int_{y_J}^\infty
 \left[f_T  y^3 \left(\frac{y^8f_T}{y_J^8 f_{TJ}} - 1 \right)  \right]^{-1/2} dy \ ,
 \label{L1}
\end{equation}
where the factor of 2 arises from adding up both sides of the ``U".
Thus, for each value of $u_J$ (or, equivalently, of $y_J$), there is a unique solution with a given value of $L\,T$ that
can be numerically computed by integrating (\ref{L1}). This is represented in figure \ref{fig_LT}.
The figure also displays some profiles for different values of $y_J$. 

The next step is to understand in which cases the U-shaped profile is energetically preferred to the disconnected brane-antibrane pair.
We have to compare the on-shell actions of both cases. Let us first express (\ref{SDBI1}) in terms of the dimensionless constants.
We write $V_{1,3}$ for the (infinite) volume of Minkowski space and $V_{S^4}$ for the volume of the internal four-sphere.
We get
\begin{equation}
S_{DBI}= 
K \int  y^{5/2} \sqrt{1 + f_T y^3 (\partial_y x)^2}   dy \ ,
\label{SDBI2}
\end{equation}
where $K=\frac{T_8}{g_s}V_{1,3}V_{S^4} R^{3/2} u_T^{7/2}$ is a constant factor, common to all brane configurations.
For the disconnected configuration, taking into account the factor of 2 for the brane-antibrane pair and inserting a UV cut-off,
\begin{equation}
 S_{DBI}|_{d} = 2 K \int_1^{y_{cut}} y^{5/2} dy  \ .
\end{equation}
For the connected configuration, we can insert the value of  $(\partial_y x)$ for the solution, as given in (\ref{yofx1}),
\begin{equation}
S_{DBI}|_c = 2K \int_{y_J}^{y_{cut}} y^{5/2} \left(1-\frac{y_J^8 f_{TJ}}{y^8 f_T} \right)^{-1/2} dy \ .
\end{equation}
We are interested in the difference $\Delta S_{DBI} =S_{DBI}|_c - S_{DBI}|_d $. This difference is not divergent and the UV cut-off can be
safely removed.
Splitting $S_{DBI}|_{d}$ into two integrals below and above $y_J$, we have
\begin{equation}
\frac{\Delta S_{DBI}}{K}= 2 \int_{y_J}^{\infty} y^{5/2} \left[\left(1-\frac{y_J^8 f_{TJ}}{y^8 f_T} \right)^{-1/2} - 1\right] dy
-  \frac47 (y_J^{7/2} -1) \ .
\label{eqDS}
\end{equation}
The value of $\Delta S_{DBI}$ can be computed numerically as a function of $y_J$.
It turns out that  $\Delta S_{DBI}>0$ for $y_J<y_{\chi SB} \approx 1.3592$, 
a case in which the disconnected configuration is preferred and chiral symmetry is preserved. 
Conversely,  $\Delta S_{DBI}<0$ for $y_J>y_{\chi SB}$ and the connected configuration is preferred.
 The value of $y_{\chi SB}$ corresponds to  $(L T)_{\chi SB}\approx 0.1538$.

\begin{figure}[htb]
\begin{center}
\includegraphics[width=\textwidth]{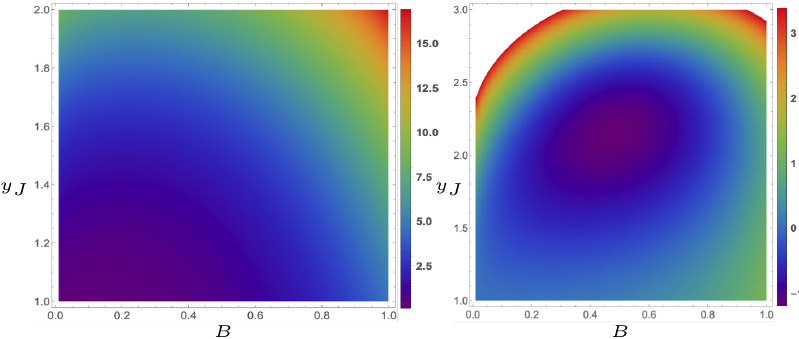}
\end{center}
\caption{Numerically computed values of
$K^{-1} \Delta  {S_{DBI}}(y_J,B)$ for two different values of $\tilde L$. On the left ($\tilde L=1$), the minimum is at $y_J=1$, $B\to 0$ and therefore the disconnected solution is preferred. On the  right ($\tilde L=0.5$), the minimum is at $y_J=2.15$, $B=0.48$, the connected solution has lower energy and chiral
symmetry breaking is to be expected.  It is interesting to
notice that the disconnected solution  ($y_J=1$, $B\to 0$)  remains a local minimum of the action in all the cases.}
\label{variat_DS}
\end{figure}

We now demonstrate that a variational approach can provide a good approximation to these results. 
Let us consider a family of profiles for a length $L$ of the form
\begin{equation}
y=y_J+ B \left[ \arctanh \left( \frac{2x}{\tilde L}\right)\right]^2 \ ,
\label{vari_profile}
\end{equation}
where $\tilde L$ is the distance between the brane and the antibrane in the coordinate $x$, which, taking (\ref{redef}) into account, is
related to $L$ as
\begin{equation}
\tilde L= \frac{4\pi }{3}L T \ .
\end{equation}
The expression (\ref{vari_profile}) can be inverted,
\begin{equation}
x=\frac{\tilde L}{2} \tanh\left(\frac{\sqrt{y-y_J}}{\sqrt{B}}\right) \ .
\label{vari_profile2}
\end{equation}
The parameters $y_J$ and $B$ are here variational constants that can take values $1\leq y_J<\infty$, $0<B<\infty$.
It is important to remark that the variational profile
 smoothly interpolates between a $U$-shaped profile and the chiral symmetry preserving profile that is recovered in the limit
$y_J = 1$, $B \to 0$.
For a particular $\tilde L$, the values of $y_J$ and $B$  have to be determined by
 minimizing the on-shell action attained after inserting (\ref{vari_profile2}) in
\begin{equation}
\frac{\Delta S_{DBI}}{K} = 2  \int_{y_J}^\infty  y^{5/2} \left[ \sqrt{1 + f_T y^3 (\partial_y x)^2} -1\right]  dy-  \frac47 (y_J^{7/2} -1) \ ,
\label{vari_acti}
\end{equation}
where we have used (\ref{SDBI2}) and subtracted the straight brane-antibrane pair.
Figure \ref{variat_DS} depicts two examples of the behavior of $\Delta S_{DBI}$ as a function of the variational parameters.

\begin{figure}[htb]
\begin{center}
\includegraphics[width=0.6\textwidth]{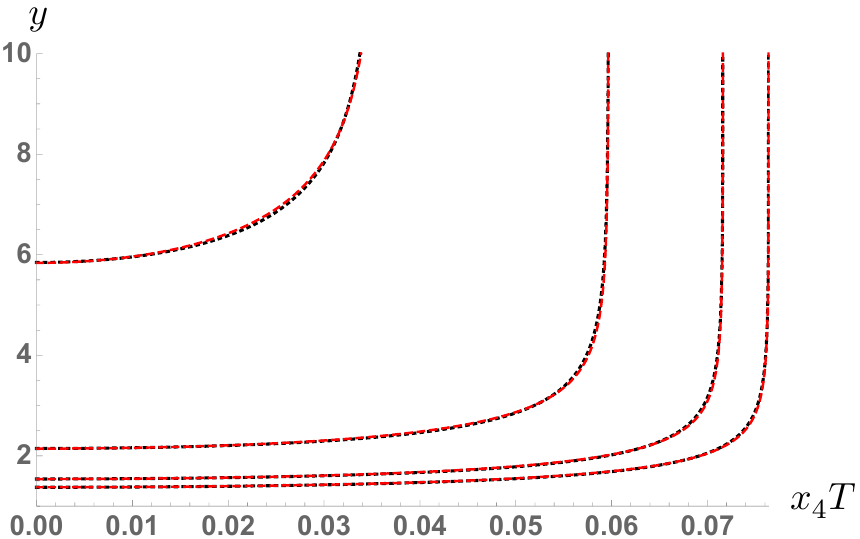}
\end{center}
\caption{Comparison of the numerical profiles (red dashed lines) with the variational profiles (black dotted lines) for four cases:
$\tilde L=0.64$, $\tilde L=0.6$, $\tilde L=0.5$ and $\tilde L=0.3$. The lines are hardly distinguishable, 
showing that the variational profile is a very good approximation
to the exact profile.}
\label{fig_compare_profile}
\end{figure}

With this procedure, a variational  approximation to the lowest energy profile can be found for any value of $\tilde L$.
Figure \ref{fig_compare_profile} shows that the approximation is quite accurate. 
To further emphasize that this variational approach captures the physics very well, we can compute the value of
$\tilde L$ at which the phase transition occurs. Numerically solving the exact equations (namely finding
from eq.~(\ref{eqDS}) the value of $y_J$ for which $\Delta S_{DBI}$ vanishes and inserting it in (\ref{L1})), we obtain
$\tilde L_{\chi SB}=0.6444$. From the variational approach, we find $\tilde L_{\chi SB}=0.6442$.
We have introduced 
this analytic approximation to the brane profiles in order to simplify the computation of vacuum decay that will be discussed below. Nevertheless,
it is natural to expect that it may also prove useful to study other properties of the WSS as, e.g., the relation between the excitations of the 
branes in the connected and disconnected phases \cite{Paredes:2008nf}.

\subsection{Flavor brane bubbles}
\label{subsec:flavor_bubbles}

We have seen that for $\tilde L = \frac{4\pi}{3} L\,T < 0.644$, the chiral symmetry breaking configuration is energetically preferred 
(it is the ``true vacuum'')
and therefore for lower  temperatures
the chirally symmetric vacuum (the ``false vacuum'') can decay through bubble nucleation \cite{Coleman:1977py,Coleman:1980aw,Linde:1980tt,Linde:1981zj}. 
The bubble would correspond to 
a ``bounce solution''. Namely, we look for a regular solution of the equations of motion obtained from the 
Euclidean action that interpolates between a configuration related to the true vacuum at the center of the bubble and the false vacuum far away from it. 
Our goal is to produce estimates for the production rate of vacuum decay bubbles. As in the deconfinement phase transition case, we will discuss 
ansatze with
$O(3)$ \cite{Coleman:1977py,Coleman:1980aw} 
and $O(4)$ \cite{Linde:1980tt,Linde:1981zj} symmetries.

\subsubsection{$O(3)$-symmetric bubbles}

We start by rewriting the metric with the Euclidean physical space in spherical coordinates, with $\rho$ as the radial coordinate,
\begin{equation}
ds_E^2= \left(\frac{u}{R}\right)^{3/2} \left[f_T(u) dt^2 +d\rho^2 + \rho^2 d\Omega_2^2 + dx_4^2\right] + \left(\frac{R}{u}\right)^{3/2} \left[\frac{du^2}{f_T(u) }+u^2 d\Omega_4^2 \right] \ .
\label{deconf_metric2}
\end{equation}
Considering an ansatz in which $x_4(u,\rho)$, the DBI action reads
\begin{equation}
S_{DBI}=\frac{T_8}{g_s}\int d^9x \rho^2 \left(\frac{u}{R}\right)^{-3/2}u^4 \sqrt{1 + f_T(u) \left(\frac{u}{R}\right)^3 (\partial_u x_4)^2 +(\partial_\rho x_4)^2}  \ .
\label{SDBI3}
\end{equation}
We can use (\ref{redef}), (\ref{redef2}) together with
\begin{equation}
\rho = \s \,  u_T^{-1/2} R^{3/2} = \s \frac{3}{4\pi T} \ ,
\end{equation}
in order to extract all the dimensionful factors from the integral.
In terms of quantities of the dual field theory, 
we find\footnote{The value of $R^3$ given in (\ref{wittenbackground}), the value of $u_T$ is given in (\ref{uTval}) and the integral over $t$ is $T^{-1}$ as implied
from (\ref{uTval}). The volumes of the two- and four-spheres are $V_{S^2}=4\pi$ and $V_{S^4}=8\pi^2/3$. The tension of the D8-brane is
$T_8=(2\pi)^{-8} l_s^{-9}$. Finally, we inserted the value of
$l_s g_s$ given in (\ref{holomaps}).}
\begin{equation}
S_{DBI}=\frac{N T^3 \lambda^3}{486 M_{KK}^3}\ \tilde S \ ,
\label{factors_action}
\end{equation}
where
\begin{equation}
\tilde S =\int \int \s ^2  y^{5/2} \sqrt{1 + (y^3-1) (\partial_y x)^2 + (\partial_\s x)^2}    d \s dy \ .
\label{generalS2}
\end{equation}
Once extracted the factor written in (\ref{factors_action}), the renormalized on-shell action is
\begin{equation}
\Delta \tilde S = 2 \int_0^\infty d \s \, \s^2 \left(  \int_{y_J(\s)}^\infty  y^{5/2} \left[ \sqrt{1 + (y^3-1)  (\partial_y x)^2+ (\partial_\s x)^2} -1\right]  dy-  \frac27 (y_J(\s)^{7/2} -1)
\right) \ ,
\label{vari_acti2}
\end{equation}
where we have subtracted the straight brane-antibrane pair.
We can derive the Euler-Lagrange equation for $x(y,\s)$ from the Lagrangian density,
\begin{equation}
\partial_y\left(\frac{\s ^2 y^{5/2}  (y^3-1) (\partial_y x)}{ \sqrt{1 + (y^3-1) (\partial_y x)^2 + (\partial_\s x)^2}}
\right)+
\partial_\s \left( \frac{\s^2 y^{5/2}  (\partial_\s x)}{ \sqrt{1 + (y^3-1) (\partial_y x)^2 + (\partial_\s x)^2}}
\right)=0 \ .
\label{fullPDE}
\end{equation}

\begin{figure}[htb!]
\begin{center}
\includegraphics[width=0.49\textwidth]{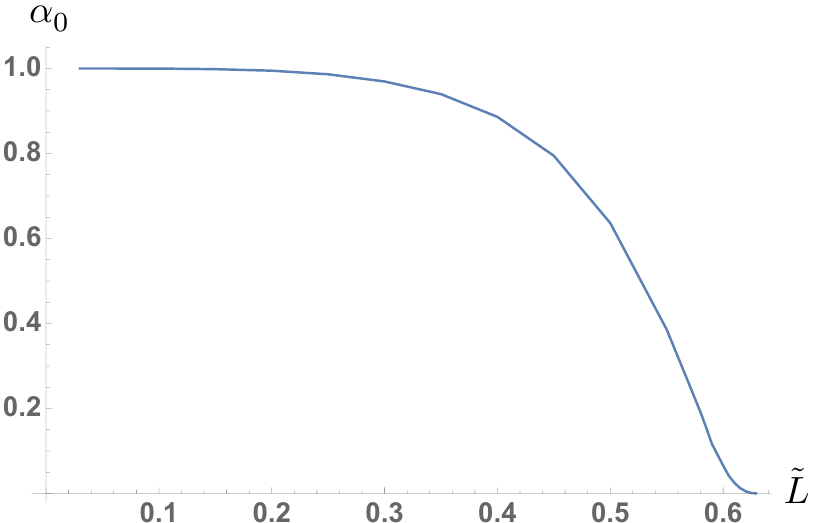}
\includegraphics[width=0.49\textwidth]{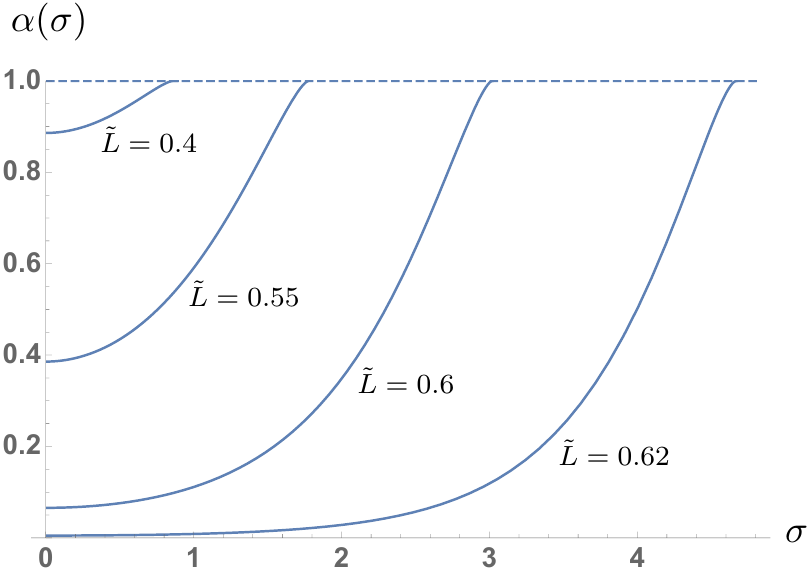}\\
\vskip.5cm
\includegraphics[width=0.49\textwidth]{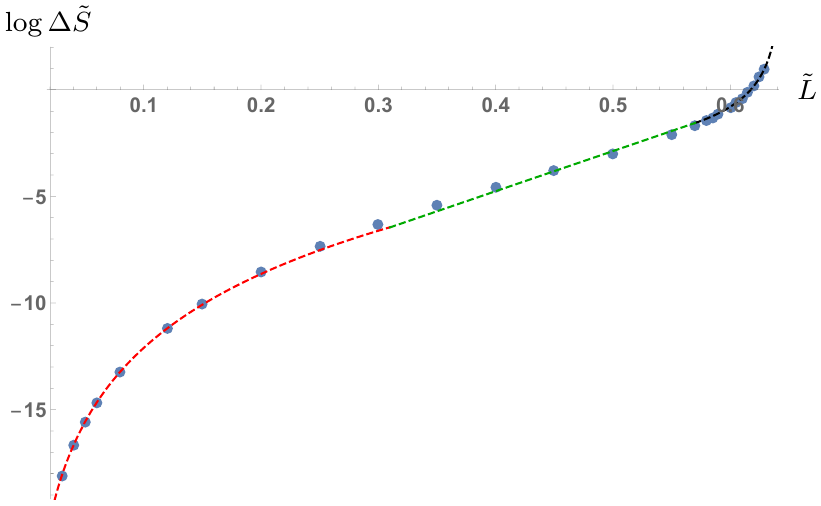}
\includegraphics[width=0.49\textwidth]{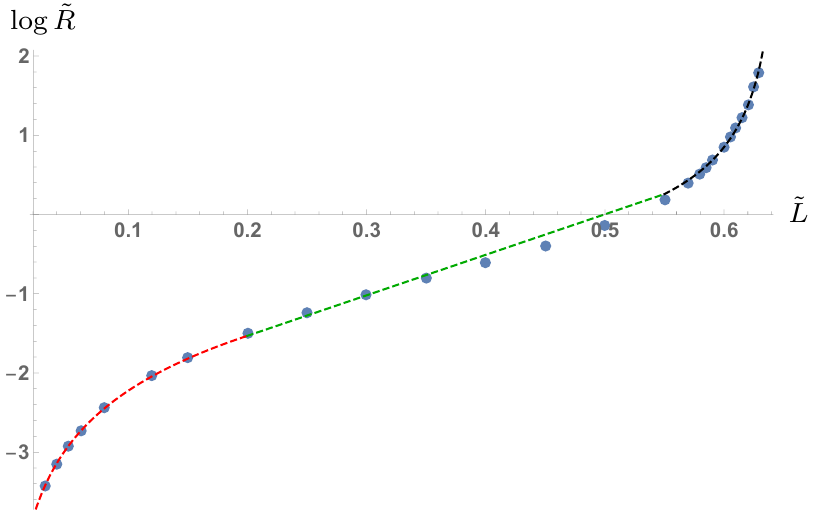}
\end{center}
\caption{On the top left, we depict the values of $\alpha_0$ in the variational approximation to the bounce solution as a function of $\tilde L$. 
Notice that $\alpha_0\to 0$ for $\tilde L \to 0.6442$ and the interior of the bubble is very close to the true vacuum. On the other hand 
$\alpha_0\to 1$ as $\tilde L \to 0$. On the top right, we depict the numerically found profiles for four values of $\tilde L$.
On the bottom, we depict the on-shell action and the radius of the $O(3)$-bubble as a function of $\tilde L$ in a semilogarithmic scale. The dots represent numerically computed data and the dashed lines correspond to the analytic approximation given in eqs.~(\ref{fits1An}) and (\ref{fits1An-R}). Different colors correspond to different expressions of the piecewise functions.}
\label{fig:alpha0}
\end{figure}

\begin{figure}[htb]
\begin{center}
\includegraphics[width=0.495\textwidth]{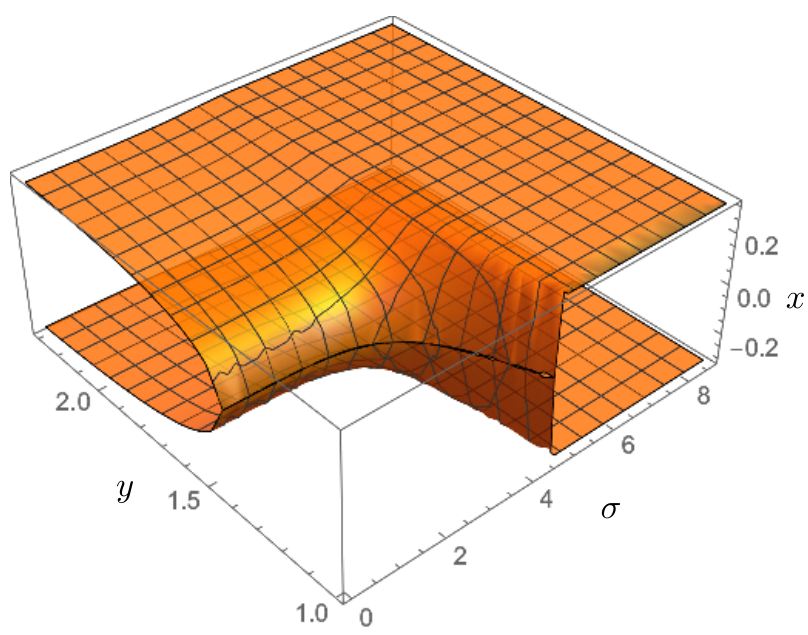}
\includegraphics[width=0.495\textwidth]{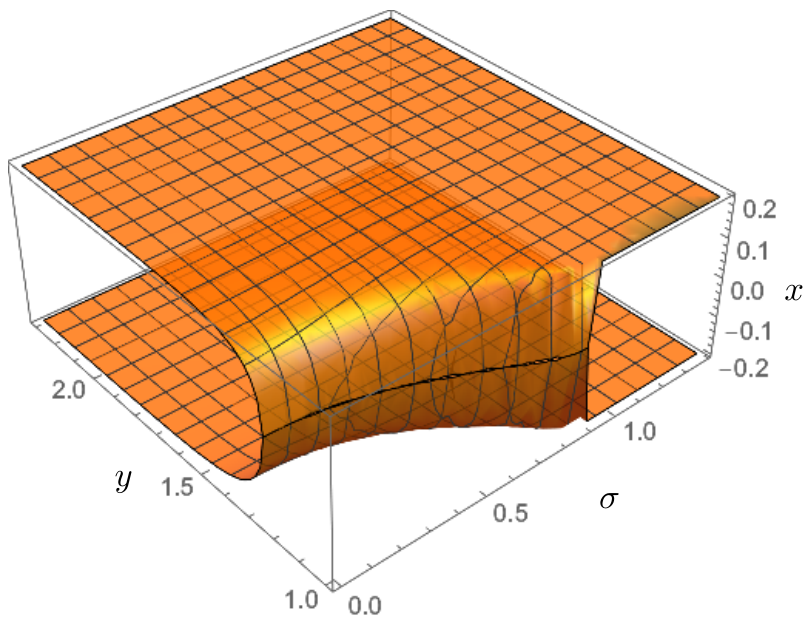}\\
\end{center}
\caption{Plots for the bounce profile $x(y,\sigma)$ in two cases:  $\tilde L = 0.62$ (left) and   $\tilde L = 0.4$ (right). 
The configurations smoothly interpolate between U-shaped profiles at $\sigma=0$ and disconnected branes at $\sigma \to \infty$.
The solution on the left can be regarded as a thin wall bubble: a U-shaped configuration very close to the true vacuum exists for a finite
range of $\sigma$ which then rapidly evolves into the false vacuum. On the other hand, the embedding on the right can be considered
as a thich wall configuration.}
\label{fig:3Dprofiles}
\end{figure}

Numerically solving (\ref{fullPDE}) is a daunting task, due to the non-linear nature of the partial differential equation. A much simpler possibility is to look for approximate solutions by using a reasonable variational
ansatz.
Taking into account the discussion of the previous section, the natural choice is to promote the $y_J$ and $B$ constants
in (\ref{vari_profile2}) to functions of $\s$, namely
\begin{equation}
x=\frac{\tilde L}{2} \tanh\left(\frac{\sqrt{y-y_J(\s)}}{\sqrt{B(\s)}}\right) \ .
\label{vari_profile3}
\end{equation}
We will use a further simplification, assuming that the bounce is a straight line in the $y_J,B$ plane. This simplifies the computations because there
is only one function of one variable that is unknown. Take
\begin{eqnarray}
y_J(\s)&=& y_{J,tv} -  (y_{J,tv}-1)\alpha(\s) \ ,\nonumber\\
B(\s)&=&B_{tv}(1-\alpha(\s)) \ ,
\label{y0Balpha}
\end{eqnarray}
where the $tv$ labels mean ``true vacuum''.
This true vacuum corresponds to $\alpha(\s)=0$ and the false vacuum to $\alpha(\s)=1$.
Therefore, we insert (\ref{vari_profile3}), (\ref{y0Balpha}) into (\ref{vari_acti2}), derive the Euler-Lagrange equation for
$\alpha(\s)$ and, in analogy with \cite{Coleman:1977py}, look for the solution that satisfies  $\alpha'(0)=0$ and $\lim_{\s\to\infty}\alpha(\s) = 1$.
The idea is simple but the procedure is somewhat tricky, so we explain it here in some detail. 
First, we change variables in order to have fixed limits in the integrals,
\begin{equation}
z=\frac{y-y_J(\s)}{B(\s)} \ .
\end{equation}
The Lagrangian can be expressed as
\begin{equation}
{\cal L}=\int_0^{\infty } F \, dz+G \ ,
\end{equation}
where
\begin{eqnarray}
F&=&2 \s^2 B(\s)   (B(\s)  z+y_J(\s))^{5/2} \left(\sqrt{1 + (y^3-1)  (\partial_y x)^2+ (\partial_\s x)^2 }-1\right) \ , \nonumber\\
G&=&-\frac{4}{7}  \s^2\left(y_J(\s)^{7/2} -1\right) \ .
\label{eqFG}
\end{eqnarray}
Notice that, once $\tilde L$ is fixed, $y_{J,tv}$ and $B_{tv}$ can be computed as detailed in section \ref{RevisitingchiSB}.
Then $F$ is a function of $z,\s,\alpha (\s),\alpha '(\s)$ and
 $G$ is a function of $\s,\alpha (\s)$.
Thus, $\partial_{\alpha'(\s)} {\cal L}= \int_0^{\infty } \partial_{\alpha'(\s)} F \, dz$ and we can write
\begin{equation}
\frac{d}{d \s}\left[\partial_{\alpha'(\s)} {\cal L}\right] \equiv  \int_0^{\infty } H \, dz + \alpha ''(\s) \int_0^{\infty } J \, dz \ ,
\end{equation}
where $H$ and $J$ depend on $z,\s,\alpha (\s),\alpha '(\s)$ but not on $ \alpha ''(\s)$. Then, the Euler-Lagrange equation for
$\alpha(\s)$ yields
\begin{equation}
 \alpha ''(\s) =\left(\int_0^{\infty } J \, dz\right)^{-1} \left[-\int_0^{\infty } H \, dz+\int_0^{\infty } \pr{\frac{\partial F}{\partial \alpha (\s)}} \, dz+\pr{\frac{\partial G}{\partial \alpha (\s)}}\right] \ .
\end{equation}
 Having this explicit expression for  $\alpha ''(\s)$, we  set up a standard explicit fourth-order Runge-Kutta integration method for the ordinary differential equation.
The initial conditions are provided near the center,
\begin{equation}
\alpha(0)=\alpha_0 \ ,\qquad\quad  \alpha'(0)=0 \ .
\end{equation}
The goal is to  determine $\alpha_0 \in (0,1)$ in order to have $\lim_{\s\to\infty}\alpha(\s) = 1$. It turns out that if $\alpha_0$ is chosen to be too small,
$\alpha(\s)$ becomes larger than 1 at some value of $\s$ and it subsequently acquires an imaginary part. On the other hand, if $\alpha_0$ is chosen to be too large,
$\alpha(\s)$ eventually starts decreasing without reaching 1. Taking these observations into account, we set up a shooting method to determine the sought value of
$\alpha_0$.  
Once the profiles are known, we can compute the value of $\Delta \tilde S$ by inserting them in (\ref{vari_acti2}). 
Figure \ref{fig:alpha0} presents some numerical results for the variational function $\alpha(\sigma)$, its value at the center of the bubble $\alpha_0$,
the on-shell action of the bounce solution and the radius of the bubble. In particular, the dimensionless radius $\tilde R$ is defined as the value of
$\sigma$ for which $\alpha$ is halfway between its value at the center and its value in the false vacuum, namely
$\alpha(\tilde R) = (\alpha_0+1)/2$.
For illustrative purposes, 
 we depict in figure \ref{fig:3Dprofiles} two examples  of the brane profiles $x(y,\sigma)$ for bounce solutions.

It is useful to have some analytic approximation for the functions $\Delta \tilde S (\tilde L)$, $\tilde R (\tilde L)$. 
We propose the following expressions, that match quite precisely the numerical results:\footnote{For values of $\tilde L$ near 0.6442, the numerics becomes very delicate and we 
have not been able to obtain reliable results for $\tilde L>0.63$. However, we assume in (\ref{fits1An}), (\ref{fits1An-R}) that $\Delta \tilde S$ 
diverges as $(0.6442\, -\tilde L)^2$ and $\tilde R$ as $(0.6442\, -\tilde L)$, as it should be expected from 
a thin wall approximation similar to  \cite{Coleman:1977py}, and find good agreement.}
\begin{equation}
\Delta \tilde S \approx \begin{cases}
0.555 \tilde L^5 \qquad\qquad\qquad\qquad\  \ (\tilde L \leq 0.31)\\
4.61 \times 10^{-6} \exp (18.8 \tilde L) \qquad\, ( 0.31\leq \tilde L  \leq 0.57)\\
\frac{0.000467}{(0.6442\, -\tilde L)^2}+\frac{0.00937}{0.6442\, -\tilde L}\qquad \q  \,(\tilde L \geq 0.57)
\end{cases}
\label{fits1An}
\end{equation}
\begin{equation}
\tilde R \approx \begin{cases}
1.081 \tilde L \q \qquad\qquad\qquad\ \,  \ (\tilde L \leq 0.2)\\
0.0777 \exp (5.11 \tilde L) \qquad \q ( 0.2\leq \tilde L  \leq 0.55)\\
\frac{0.0872}{(0.6442\, -\tilde L)}+0.369 \qquad \q \, \,(\tilde L \geq 0.55)
\end{cases}
\label{fits1An-R}
\end{equation}

\subsubsection{$O(4)$-symmetric bubbles}

When the radius of the bubble is much smaller than the inverse of the temperature, one expects to have bubbles with $O(4)$-symmetry in 
the Euclidean spacetime \cite{Linde:1980tt,Linde:1981zj}. However, the blackening factor $f_T(u)$ in (\ref{wittenbackground}) breaks
the $O(4)$-symmetry and an ansatz of the form $x_4(u,\rho)$ where $\rho$ is a radial coordinate in the $t-x_i$ four-dimensional
space is not consistent with the equations of motion. 
Still, it is natural to expect bubble solutions with non-trivial behavior along the
time coordinate, for instance with an ansatz of the type $x_4(u,\rho,t)$. Solving the problem with this ansatz, either integrating the exact
equation or with a reliable approximation seems extremely difficult and is beyond the scope of the present work. Nevertheless, we can
get an order of magnitude estimate by considering a ``naive $O(4)$ configuration" in which we just neglect the $O(4)$ breaking due to the blackening factor.\footnote{Notice that the results will produce an underestimation of the action since the presence of the blackening factor tends to increase it.}
As discussed in section \ref{effectiveactionsection}, we do this by simply considering the measure $d ^4 x$ to be given by $d \O_3 d \r \r ^3$, where $d \O_3$ is the measure of the three-sphere. By changing accordingly (\ref{vari_acti2}) and (\ref{eqFG}), we can follow the steps explained in the previous section and find the following approximate expression for the on-shell action:
\begin{equation}
\Delta \tilde S \approx \begin{cases}
0.638 \tilde L^6  \qquad\qquad\qquad\qquad\ \  (\tilde L \leq 0.22)\\
3.91 \times 10^{-7} \exp (23.8 \tilde L) \qquad \, ( 0.22\leq \tilde L  \leq 0.54)\\
\frac{0.0000432}{(0.6442\, -\tilde L)^3}+\frac{0.00118}{(0.6442\, -\tilde L)^2} \qquad  \ (\tilde L \geq 0.54)
\end{cases}
\label{fits2An}
\end{equation}

\begin{figure}[htb]
\begin{center}
\includegraphics[width=0.495\textwidth]{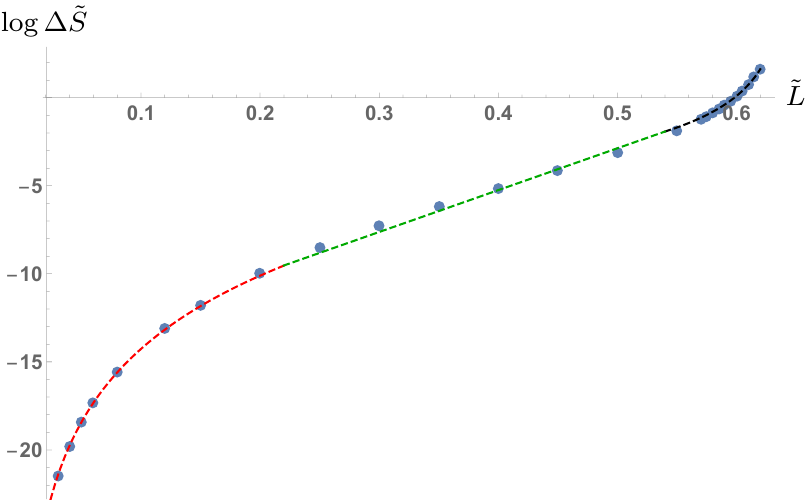}
\includegraphics[width=0.495\textwidth]{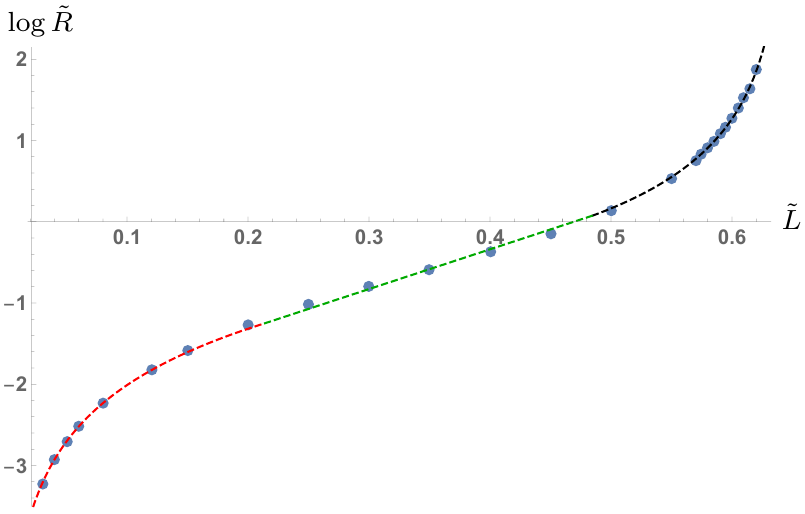}
\end{center}
\caption{The on-shell action and the radius of the $O(4)$-bubble as a function of $\tilde L$ in a semilogarithmic scale. The dots represent numerically computed data
and the dashed lines correspond to the analytic approximation given in eqs.~(\ref{fits2An}) and (\ref{fits3An}). Different colors correspond to different expressions of the piecewise functions.}
\label{fig:O4}
\end{figure}

We can also study the radius of the bubble. Defining $\tilde R$ as above, we find the approximate expressions

\begin{equation}
\tilde R \approx \begin{cases}
1.34 \tilde L \qquad\qquad\qquad  \  \ \ (\tilde L \leq 0.21)\\
0.101 \exp (4.89 \tilde L) \qquad\, ( 0.21\leq \tilde L  \leq 0.49)\\
\frac{0.151}{(0.6442\, -\tilde L)}+0.131 \qquad (\tilde L \geq 0.49)
\end{cases}
\label{fits3An}
\end{equation}
Figure \ref{fig:O4} depicts some numerical results compared to their fits given in eqs.~(\ref{fits2An}), (\ref{fits3An}).

It is important to recall that the $O(4)$ configuration could start playing a role only if the bubble radius is smaller than the radius of the time circle. It is easy to verify that this condition can be satisfied only for $\tilde L\lesssim 0.386$.

%%%%%%%%%%%%%%%%%%%%%%%%%%%%%%%%%%%%%%%%
\subsection{Bubble nucleation rate}
In principle, the rates for the bubble nucleations are provided by formula (\ref{Gamma})
\bea \label{Gamma2}
\Gamma & = & {\rm Max}\left[T^4 \left( \frac{S_{3,B}}{2\pi T} \right)^{3/2} e^{-S_{3,B}/T}      ,  \left( \frac{S_{4,B}}{2\pi \rho_w^2}  \right)^2  e^{-S_{4,B}}   \right]   \nonumber \\
& = & M_{KK}^4 {\rm Max}\left[ \left(\frac{\tilde T  \bar f_{\chi}^{2/3}}{0.35 \lambda^{1/3} N^{1/3}}\right)^4 \left( \frac{S_{3,B}}{2\pi T} \right)^{3/2} e^{-S_{3,B}/T}      ,  \left( \frac{S_{4,B}}{2\pi \bar\rho_w^2}  \right)^2  e^{-S_{4,B}}   \right] \,,      
\eea
where we have introduced the dimensionless quantities 
\begin{subequations}
\ba
\tilde T &\equiv& \frac{T L}{0.1538} \simeq 0.35 (\lambda N)^{1/3} \frac{T}{M_{KK}^{1/3} f_{\chi}^{2/3}}\ , \\ 
\bar f_{\chi} &\equiv& \frac{f_{\chi}}{M_{KK}}\,, \qquad \bar \rho_w \equiv \rho_w M_{KK} \simeq 0.35 (\l N)^{1/3} \frac{3\tilde R}{4\pi \tilde T \bar f_{\chi}^{2/3}}\ ,
\ea
\end{subequations}
so that the critical temperature for the chiral symmetry breaking transition corresponds to $\tilde T=1$ and the chiral symmetry breaking scale is given, as a function of the asymptotic brane separation $L$, by \cite{Aharony:2006da, Bigazzi:2019eks}\footnote{Note that in this paper a different convention on the coupling w.r.t.~\cite{Bigazzi:2019eks} is used: $\lambda_{here}=2\lambda_{there}$.} 
\be
f_{\chi}^2 \simeq 0.1534 \frac{\lambda N}{32 \pi^3} \frac{1}{M_{KK} L^3}\ .
\ee 
As we have outlined before, the symmetries of the black hole background do not allow for (simple) $O(4)$ solutions, so that the analysis of the previous subsection can, at best, be considered as providing a rough estimate of some limiting value of the corresponding bounce action. Hence, here, we will just focus on the $O(3)$ bounce. 

\begin{figure}
\center
\includegraphics[scale=1]{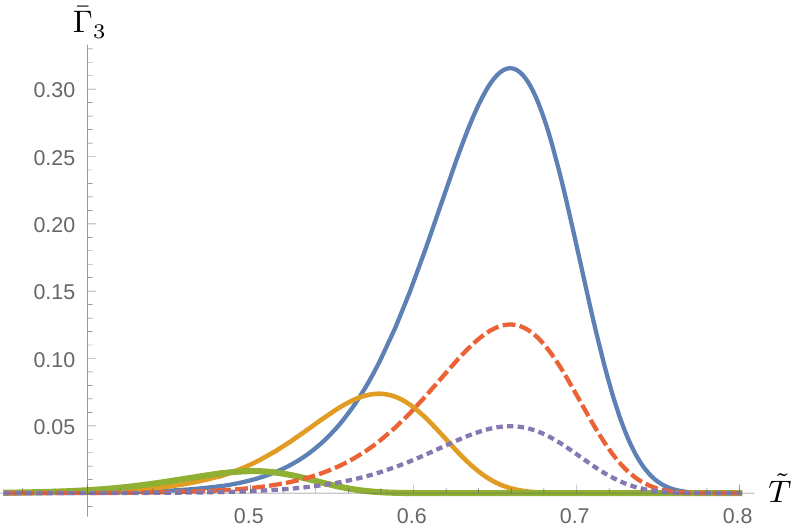}\includegraphics[scale=1]{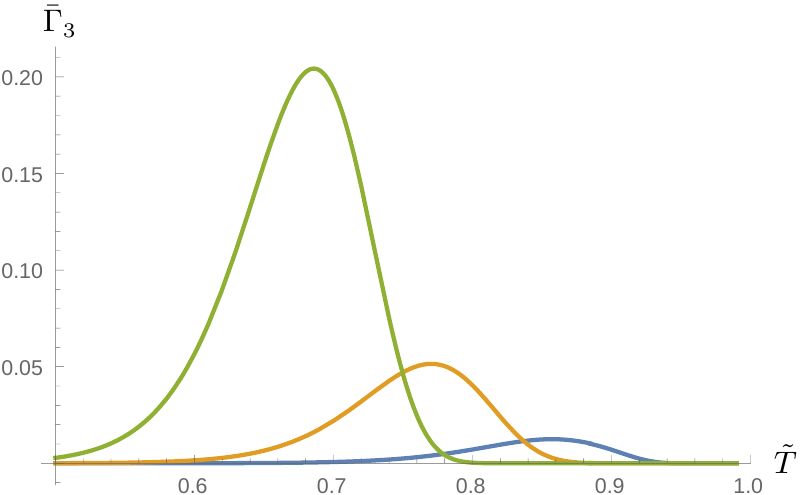}
\caption{Plots of $\bar \Gamma_3 \equiv \Gamma_3/M_{KK}^4$ for different values of parameters. On the left, the rate magnitude is quenched as $N$ is increased (solid blue, dashed and dotted lines correspond to $N=10, 20, 40$ with $\lambda=10, \bar f_{\chi}= 10$) and as $\lambda$ is increased (solid blue, orange, green lines correspond to $\lambda=10, 20, 40$ with $N=10, \bar f_{\chi}= 10$). On the right, the magnitude increases as $\bar f_{\chi}$ is increased (blue, orange and green lines correspond to $\bar f_{\chi}= 2, 4, 8$ with $\lambda=10, N= 10$).}
 \label{figratesO31}
\end{figure}

The rate for the $O(3)$ bubble depends on three distinct parameters: $\lambda, N$ and $\bar f_{\chi}$.
Its behavior when these parameters are separately varied is shown in figure \ref{figratesO31}.
Increasing $\lambda$ both quenches the rate and shifts the peak to smaller temperatures while increasing $N$ has essentially only a quenching effect. Instead, the rate magnitude is enhanced if the chiral symmetry breaking  scale $\bar f_{\chi}$ is increased, while the peak is shifted to smaller temperatures.

%%%%%%%%%%%%%%%%%%%%%%%%%%%%
\section{Conclusions}
In this paper, we have studied the dynamics of first-order phase transitions in strongly coupled planar gauge theories. Using the holographic correspondence as a tool we have been able to compute the decay rate of the false vacuum which proceeds through the nucleation of bubbles in the metastable phase. As discussed in the seminal papers \cite{Coleman:1977py,Coleman:1980aw,Linde:1980tt,Linde:1981zj}, the decay probability per unit time and unit volume in the semiclassical limit is given by $\Gamma=A e^{-S_B}$, where $A$ is a certain functional determinant which is often approximated using dimensional analysis, and $S_B$ is the on-shell action for the bounce.

In holographic models like those examined in this paper, the first-order phase transition can be related, in the dual picture, either to a change of the gravity background (a Hawking-Page transition for instance) or to a change of the embedding of some probe brane on a fixed background. 

The first case is precisely what arises when the dual quantum gauge theory experiences a first-order confinement/deconfinement transition. Describing the dynamics of the transition in the gravity side requires developing an off-shell formalism which may allow to follow the jump from a black hole solution describing the deconfined phase to a ``solitonic'' solution describing the confined one. Deriving the complete solution for the mixed fluctuations of the metric and the other background fields would be a daunting task, thus we have adopted a simplified practical approach, introduced in \cite{Creminelli:2001th} for Randall-Sundrum models with an AdS$_5$ dual description. 

The approximation consists in modeling bubble dynamics by means of an effective action for a single scalar field. This field was actually a parameter in the original homogeneous gravity solutions related to the two phases: it was the horizon radius in the black hole case and the minimum of the holographic radial coordinate in the solitonic background. In the effective off-shell Euclidean description, these parameters are combined into a space-dependent field $\Phi(\rho)$ where $\rho^2= t^2 + x_i x_i$ or $\rho^2= x_i x_i$ (with $x_i$ being the 3d space coordinates) depending on the symmetry of the bubbles. At low temperatures, where the vacuum decay is mostly driven by quantum tunneling, the bubble is expected to have an $O(4)$ symmetry. At large temperatures, where thermal fluctuation dominates, the bubble should have instead an $O(3)$ symmetry.

We have started this paper by revisiting the compact Randall-Sundrum model examined in \cite{Creminelli:2001th}. In this seminal paper, and in the following literature, a missing piece in the analysis of the bounce action in the deconfined phase (dual to an AdS$_5$ black hole) was the kinetic term for the field $T_h(\rho)$ related to the horizon radius. Using holographic renormalization we have been able to compute this term.

Holographic renormalization has also been the relevant tool we have adopted in studying the dynamics of the confinement/deconfinement transition in the top-down Witten-Sakai-Sugimoto (WSS) model. To the best of our knowledge, this is the first time the phase transition dynamics is studied in a full-fledged top-down holographic model. We have been able to extract the effective bounce action and to compute the bubble nucleation rate as a function of the model parameters. Analytic expressions have been also provided in the thick and thin wall approximations. 

The second kind of transition we have examined is the very special chiral symmetry breaking/restoration one which, provided certain parameters of the WSS model are opportunely tuned, occurs in the deconfined phase, with a critical temperature which is larger than the one for deconfinement. In this case, the two phases are related to two different solutions for the embedding of D8-brane probes in the black hole background describing the deconfined phase. The off-shell description of the transition consists in promoting the embedding function (which is originally dependent only on the holographic radial direction) to a $\rho$-dependent field. 
What is relevant in this case is that in principle the DBI action for the branes is enough to deduce the on-shell action for this field. 
However, the non-linearities inherent to the DBI action render the complete analysis very challenging. We have been able to tackle the problem by using a powerful variational approach which could hopefully be useful for treating more general (static and dynamical) problems related to flavor-brane dynamics in WSS and similar models. Again this has allowed us to compute the bubble interpolating between the two configurations and the nucleation rate.

It would be interesting to apply the techniques employed in this paper to study other holographic first-order transitions, for example involving finite density states.

Our analysis has been in part motivated by the exciting perspective, offered by near-future experiments, to detect signals of possible cosmological first-order phase transitions which could have occurred in the early Universe, as predicted in many beyond the Standard Model scenarios. Bubble nucleation, expansion and collision, and further collective dynamics of the underlying plasma are expected to be the source for a stochastic gravitational wave (GW) background which, depending on the amount of energy released after the transition, could have a power spectrum entering the sensitivity regime of future ground-based and space-based experiments. Predicting the power spectrum from first principles requires precisely to compute the relevant parameters describing the dynamics of the phase transition. Our analysis provides the tools to compute an approximation of the complete set of these parameters for the case of the WSS model. If the latter is used to describe the strongly coupled dynamics of some hidden sector, then our analysis would allow to provide falsifiable predictions on the GW signals. The advantage of using a top-down holographic model would be that the various approximations which are made for deducing the relevant parameters would be perfectly under control. We will devote a forthcoming paper \cite{draftph} to this very fascinating subject.

%%%%%%%%%%%%%%%%%%%%%%%%%%%%%%%%
\vskip 15pt \centerline{\bf Acknowledgments} \vskip 10pt 

\noindent 
We are indebted to Riccardo Argurio, Chiara Caprini, Paolo Creminelli, Luigi Delle Rose, Alberto Mariotti, Alberto Nicolis and Diego Redigolo for invaluable insights. A.L.C. would like to thank the Galileo Galilei Institute for theoretical physics and the ULB for their hospitality during the preparation of this work. A.C. thanks the ULB for the kind hospitality during the preparation of this work.

%%%%%%%%%%%%%%%%%%%%%%%%%%%%%%%%%%%%%%%%%%%%%%%%%%%%%%%%%%

\appendix
 %%%%%%%%%%%%%%%%%%%%%%%%%%%%%%%%%%%%%%%%%%%%%
\section{Thick and thin wall approximations for the deconfinement transition}
\label{appthickthin}
In this appendix, we provide some analytical estimates of the bounce action, of the radius of the bubbles and of the vacuum decay rate related to the confinement/deconfinement phase transition. We adopt the two standard thick and thin wall approximations.
%%%%%%%%

\subsection{The $O(4)$ bubble}

We follow the procedure discussed in \cite{Nardini:2007me}. Let us assume that the nucleation temperature is much smaller than $T_c$.
In such a regime, if the bubble radius is smaller than $1/(2\pi T)$, the system has $O(4)$ symmetry and its physics can be captured by the thick wall approximation. We recall that, in our setup, the Euclidean action with $O(4)$ symmetry reads
\be
S_4 (\Phi) = \frac{8\pi^4 g}{3^5} \int_{0}^{\infty}d\bar\rho\,\bar\rho^3 \left[ a\, \Phi'^2 + \Theta(\Phi) V_c(\Phi) + \Theta(-\Phi) V_d(\Phi) \right] \ ,
\ee
where
\be
a= 5-\frac{\pi}{2\sqrt{3}}
\ee
and
\bea
V_c(\Phi) &=& \frac{16\pi^2}{9}\left(5\Phi^3-\frac{3}{\pi}\Phi^{5/2}\right) \ ,\nonumber \\
V_d(\Phi) &=& -\frac{16\pi^2}{9}\left(5\Phi^3+\frac{3}{\pi}\bar T (-\Phi)^{5/2}\right) \ .
\eea
The total potential has a false vacuum at $\Phi=\Phi_d = -{\bar T}^2/(4\pi^2) $ where $V_d= - {\bar T}^6/(36\pi^4)$ and a true vacuum at $\Phi=\Phi_c=1/(4\pi^2)$ where $V_c=-1/(36\pi^4)$. 

The $O(4)$ bounce is a solution $\Phi _B$ of the equations of motion following from $S_4$ with boundary conditions 
$\Phi' _B(\bar\rho=0) =0$ and $\Phi _B(\bar\rho\rightarrow\infty)=\Phi_d$. Let us indicate by $\Phi_0$ the value of the solution at the center of the bubble (i.e.~at $\bar\rho=0$). 

Let us consider a bubble of true vacuum and (dimensionless) radius $\bar\rho_w$ nucleated in the false vacuum. What we need is the on-shell value of the action $S_4$ on the bounce solution, or, more precisely, the difference between the latter and the action computed on the false vacuum,
\be
S_{4,B}  = S_4 (\Phi _B)- \frac{8\pi^4 g}{3^5} \int_{0}^{\infty}d\bar\rho\,\bar\rho^3 V_d (\Phi_d) \ .
\label{rens4}
\ee
More explicitly, it reads
\be
S_{4,B}   =\frac{8\pi^4 g}{3^5}\left[ \int_{0}^{\infty}d\bar\rho\,\bar\rho^3 [ a\, \Phi _B '^2 - V_d (\Phi_d)]+\int_{0}^{\bar\rho_w}d\bar\rho\,\bar\rho^3 V_c(\Phi _B) + \int_{\bar\rho_w}^{\infty}d\bar\rho\,\bar\rho^3 V_d(\Phi _B)\right]\,.
\ee
If $\bar\rho_w\rightarrow\infty$, we can approximate the above expression as
\be
S_{4,B} \approx  \frac{8\pi^4 g}{3^5} \int_{0}^{\infty}d\bar\rho\,\bar\rho^3 \left[ a\, \Phi_B '^2 + V_c(\Phi _B) -V_d(\Phi_d) \right]\ .
\ee
Just as in \cite{Nardini:2007me}, let us roughly estimate this action as
\be
S_{4,B} \approx \frac{8\pi^4 g}{3^5} \left[ \bar\rho_w^3 a \left(\frac{\delta\Phi _B}{\delta\bar\rho_w}\right)^2 \delta\bar\rho_w + \frac14 \left( V_c(\Phi_0) - V_d (\Phi_d) \right) \bar\rho_w^4 \right]\ ,
\label{s4renap}
\ee
where $\delta\Phi _B= \Phi _B(0)-\Phi _B(\infty) = \Phi_0-\Phi_d$.  In the thick wall approximation
\be
\delta\bar\rho_w \approx \bar\rho_w \ ,
\ee
so that, extremizing (\ref{s4renap}) w.r.t.~$\bar\rho_w$ we find the critical bubble radius squared
\be
\bar\rho_w^2 \approx -\frac{2 a (\delta\Phi _B)^2}{ [ V_c(\Phi_0) - V_d(\Phi_d) ]} \ .
\label{rhowexp}
\ee
Now, numerical analysis shows that $\Phi_0 \approx c_0 {\bar T}^2$ at small $\bar T$ so that
\be
\delta\Phi _B = \Phi_0-\Phi_d \approx \left(c_0 + \frac{1}{4\pi^2}\right) {\bar T}^2 
\ee
and
\be
V_c(\Phi_0) - V_d(\Phi_d)= \frac{16\pi^2}{9}\left(5\Phi_0^3-\frac{3}{\pi}\Phi_0^{5/2}\right)+\frac{{\bar T}^6}{36\pi^4}\approx -\frac{16\pi}{3}c_0^{5/2} {\bar T}^5 \ .
\ee
Hence, from (\ref{rhowexp}), we get
\be
\bar\rho_w^2 \approx \frac{ 3 a }{16 \pi c_0^{5/2}} \left(c_0 + \frac{1}{4\pi^2}\right)^2 \frac{1}{\bar T}\equiv \frac{b^2}{\bar T}\ .
\ee 
Thus, the bubble radius goes like $\bar\rho_w\sim {\bar T}^{-1/2}$ when $\bar T\ll1$: this relation qualitatively reproduces what we have obtained numerically in the small $\bar T$ regime.

Recalling that $\bar\rho\equiv M_{KK}\rho$ and $M_{KK} \bar T = 2\pi T$, the above results imply that the dimensionful bubble radius in the small temperature regime scales like
\be
\rho_w \approx \frac{b}{\sqrt{2\pi\, T\, M_{KK}}} \ .
\ee
Now, an important question regarding our holographic model is whether in the limit of small enough bubble radius a $O(5)$ symmetric bubble should be used instead of the $O(4)$ symmetric one. This should be unavoidable if the bubble radius turns out to be smaller than $1/(2\pi T)$ (the length of the radius of the time circle) and, at the same time, smaller than $1/M_{KK}$ (the length of the radius of the $x_4$ circle). Let us study whether these two conditions are mutually compatible in the regime where the approximations used since now hold. The first condition implies
\be
\rho_w\ll \frac{1}{2\pi T} \q \text{hence}\q  T \ll \frac{M_{KK}}{2\pi b^2} \ ,
\ee
while the second one implies
\be
\rho_w\ll \frac{1}{M_{KK}} \q \text{hence}\q  T \gg b^2\frac{M_{KK}}{2\pi} \ .
\ee
At least parametrically, the two above conditions are not mutually compatible. Hence we argue that in the regime of parameters where the bubble is $O(4)$ symmetric, an $O(5)$ configuration cannot be consistent.
The very same considerations can be done for the directions along the four-sphere of the background.

Let us now try to see whether, in the thick wall approximation, it is possible to deduce some qualitative information about the nucleation rate. For this aim, it is enough to notice that the action (\ref{s4renap}) at the critical radius (\ref{rhowexp}) reads
\be
S_{4,B}  \approx -\frac{2 \pi^4 g}{3^5} \bar\rho_w^4 [ V_c(\Phi_0) - V_d(\Phi_d) ] \approx c_4\, g\, {\bar T}^3 \ .
\label{s4crit}
\ee
From the fit of numerical data and the previous relations we get
\be
c_4\approx 0.39\ , \q \quad b\approx 6.6 \ .
\ee
The nucleation rate is given by
\be
\Gamma_4  =  M_{KK}^4    \frac{c_4 ^2}{\left(2\pi   \right)^2 b^4}  g ^2  {\bar T}^8   e^{-c_4\, g\, {\bar T}^3}   \ .
\ee
%%%%%%%%%%%%%%%%%%%%%
\subsection{The $O(3)$ bubble}
As explained in section \ref{effectiveactionsection},
the radius of the $O(4)$ bubble is much smaller than the dimensionless parameter $1/\bar T$ only for very small $\bar T$, i.e. $\bar T\lesssim 0.06$. Hence the use of the $O(4)$ symmetric bounce for larger values of $\bar T$ is questionable and it should be replaced by the $O(3)$ symmetric one.

The $O(3)$ bounce arises as a solution of the action $S=S_3(T)/T$ where $S_3$ is the Euclidean action with $O(3)$ symmetry, 
\be
\frac{S_3 (\Phi)}{T}=\frac{32\pi^4 g}{3^5 \bar T} \int_{0}^{\infty}d\bar\rho \bar\rho^2 \left [ a \Phi'^2 +  \Theta(\Phi) V_c(\Phi) + \Theta(-\Phi) V_d(\Phi) \right]\,.
\label{s3sut}
\ee
As already mentioned, we need the difference between the on-shell action on the bounce solution and the action evaluated on the false vacuum configuration,
\be
\frac{S_{3,B} }{T} = \frac{S_{3} (\Phi_B)}{T} - \frac{32\pi^4 g}{3^5 \bar T} \int_{0}^{\infty}d\bar\rho\,\bar\rho^2 V_d (\Phi_d) \ .
\label{rens3B}
\ee
Explicitly,
\be
\frac{S_{3,B} }{T}=\frac{32\pi^4 g}{3^5 \bar T} \left[ \int_{0}^{\infty}d\bar\rho\,\bar\rho^2 [ a\, \Phi _B '^2 - V_d (\Phi_d)]+\int_{0}^{\bar\rho_w}d\bar\rho\,\bar\rho^2 V_c(\Phi _B) + \int_{\bar\rho_w}^{\infty}d\bar\rho\,\bar\rho^2 V_d(\Phi _B)\right]\,.
\label{s3sutB}
\ee

%%%%%%%%%%%%%%%%%%%%%%%%%%%%%%%%%%%%%%%%%%%%%%
\subsubsection{Small temperatures}
\label{O3smallt}
It is worth to consider the case in which for some range of values of $\bar T\ll1$ the $O(3)$ configuration is the relevant one. In this case, we could try to use the thick wall approximation. 

In this approximation, following the same steps described in the previous subsection and using the fact that $\Phi_0\sim \bar T^2$ for small $\bar T$, we find the dimensionless bubble radius
\be
\bar\rho_w^2\approx  -\frac{a (\delta\Phi _B)^2}{[ V_c(\Phi_0) - V_d(\Phi_d) ]}\approx \frac{\tilde{b}^2}{\bar T}\ ,
\ee
for some constant $\tilde b$. The action at the critical radius above reads
\be
\frac{S_{3,B} }{T} \approx -\frac{64\pi^4 g}{3^6 \bar T} [V_c(\Phi_0) - V_d(\Phi_d)] \bar\rho_w^3 \approx c_3\, g\, {\bar T}^{5/2}\,.
\ee
The $S_4$ action is parametrically smaller than $S_3/T$. 
From the fit of numerical data and the previous relations we get
\be
c_3\approx 0.32\ , \qquad \tilde b \approx 9.3 \ .
\ee
When the $O(3)$ configuration dominates, the nucleation rate is given by
\be
\Gamma_3 = M_{KK}^4 \frac{c_3 ^{3/2}}{(2\pi)^{11/2}}  g ^{3/2} \bar T ^{31/4}   e^{-c_3\, g\, {\bar T}^{5/2}}\ .
\label{rateo3}
\ee

%%%%%%%%%%%%%%%%%%%%%%%%
\subsubsection{Large temperatures}
At large enough temperatures, the $O(3)$ configuration is definitely the dominant one. We can try to get some intuition about its physical properties using the thin wall approximation, which is expected to be valid around $T_c$, i.e.~in the $\bar T\rightarrow 1$ limit \cite{Coleman:1977py}.

In the thin wall approximation, the bounce action can be estimated as
\be
\frac{S_{3,B} }{T} \approx \frac{32\pi^4 g}{3^5 \bar T} \left[ \frac{\bar \rho_w^3}{3}\Delta V + \bar \rho_w^2 S_1\right]\ ,
\label{s3app}
\ee
where $S_1\approx S_1(T_c)$ is the bubble surface tension
\be
S_1 = 2\sqrt{a}\int_{\Phi_d}^{\Phi_c}d\Phi \sqrt{\frac{16\pi^2}{9}\left(5|\Phi|^3 -\frac{3}{\pi}|\Phi|^{5/2}\right)+\frac{1}{36\pi^4}}\approx 0.0023\ ,
\ee
and 
\be
\Delta V  =  V_c(\Phi_c) - V_d(\Phi_d)  = - \frac{1}{36\pi^4} ( 1- \bar T^6 )\ .
\ee
Extremizing the action above, we get the critical bubble radius
\be
\bar\rho_w \approx -\frac {2 S_1}{\Delta V} \approx \frac{16}{1-\bar T^6}\ .
\label{rhowthin}
\ee
This is increasing for $\bar T\rightarrow1$, in qualitative agreement with our numerical results.

In the $\bar T\rightarrow1$ limit, the action (\ref{s3app}) at the critical radius (\ref{rhowthin}) goes like
\be
\frac{S_{3,B} }{T} \approx \frac{\tilde c_3 g}{\bar T (1-\bar T^6)^2}\ , \qquad \tilde c_3 \approx 2.6 \ ,
\ee
so that in the same limit the nucleation rate (\ref{rateo3}) goes as
\be
\Gamma_3 \approx \frac{M_{KK}^4}{(2\pi)^4} \frac{\tilde c_3^{3/2}}{(2\pi)^{3/2}}\frac{g^{3/2}}{ \bar T ^{3/2}(1-\bar T^6)^3} e^{- \frac{\tilde c_3 g}{ \bar T (1-\bar T^6)^2}}\ .
\label{gammathin}
\ee 
%%%%%%%%%%%%%%%%%%%%%%%%%%%%%%%%%%%%%%%%%%%%%%%%%

\end{document}